\documentclass[preprint,12pt]{elsarticle}
\usepackage{amsmath,amsfonts}
\usepackage{algorithmic}
\usepackage{algorithm}
\usepackage{array}
\usepackage[caption=false,font=normalsize,labelfont=sf,textfont=sf]{subfig}
\usepackage{textcomp}
\usepackage{stfloats}
\usepackage{url}
\usepackage{verbatim}
\usepackage{graphicx}
\usepackage{amsmath,amssymb,amsfonts}
\usepackage{algorithmic}
\usepackage{graphicx}
\usepackage{textcomp}
\usepackage{bm}

\usepackage{amsthm}

\theoremstyle{definition}
\newtheorem{definition}{Definition}
\theoremstyle{proposition}

\newcommand{\norm}[1]{\left\lVert#1\right\rVert}

\journal{ISA Transactions}
\begin{document}
	
\begin{frontmatter}
	
\title{Does Optimal Control Always Benefit from Better Prediction? An Analysis Framework for Predictive Optimal Control}
	
\author[1]{Xiangrui Zeng\corref{cor1}}
\ead{zeng@hust.edu.cn}

\author[2]{Cheng Yin}
\ead{yinchenghust@hust.edu.cn}

\author[1]{Zhouping Yin}
\ead{yinzhp@hust.edu.cn}

\address[1]{State Key Laboratory of Intelligent Manufacturing Equipment and Technology, Huazhong University of Science and Technology, Wuhan, Hubei, China}
\address[2]{The China-EU Institute for Clean and Renewable Energy, Huazhong University of Science and Technology, Wuhan, Hubei, China}
\cortext[cor1]{Corresponding author}



\begin{abstract}
	The ``prediction + optimal control'' scheme has shown good performance in many applications of automotive, traffic, robot, and building control.
	In practice, the prediction results are simply considered correct in the optimal control design process.
	However, in reality, these predictions may never be perfect. 
	Under a conventional stochastic optimal control formulation, it is difficult to answer questions like ``what if the predictions are wrong''.
	This paper presents an analysis framework for predictive optimal control where the subjective belief about the future is no longer considered perfect.
	A novel concept called the hidden prediction state is proposed to establish connections among the predictors, the subjective beliefs, the control policies and the objective control performance.
	Based on this framework, the predictor evaluation problem is analyzed.
	Three commonly-used predictor evaluation measures, including the mean squared error, the regret and the log-likelihood, are considered.
	It is shown that neither using the mean square error nor using the likelihood can guarantee a monotonic relationship between the predictor error and the optimal control cost.
	To guarantee control cost improvement, it is suggested the predictor should be evaluated with the control performance, e.g., using the optimal control cost or the regret to evaluate predictors.
	Numerical examples and examples from automotive applications with real-world driving data are provided to illustrate the ideas and the results.
\end{abstract}

\begin{keyword}
	Model predictive control, optimal control, data-based control
\end{keyword}
\end{frontmatter}
\section{Introduction}
\label{sec:intro}

In many automotive control\cite{zengParallelHybridElectric2015a}\cite{wanProbabilisticAnticipationControl2019}, traffic control \cite{YeAsurveyofmodel2019}, robot control \cite{fridovich-keilConfidenceawareMotionPrediction2020} and building control \cite{drgonaAllYouNeed2020} applications, optimal control decisions need to be made in the presence of an uncertain future.
This uncertain future is usually caused by complicated human behaviors or highly complex environment systems, and it can have a relative large impact on the control system performance\cite{9922564}.
The control policies in these applications need to be adjusted according to different potential future scenarios\cite{8248668}.
A common way to handle this is to use a predictor to forecast the future, and then apply the optimal controller with respect to this forecasted future\cite{509370}. 
The ``prediction + optimal control'' scheme has shown good performance in practice. 
There have also been theoretical results showing that predictions for certain control problems are beneficial\cite{yuPowerPredictionsOnline2020}. 
In this paper, we refer to this type of control method as \textit{predictive optimal control}.

Predictive optimal control is closely related to model predictive control (MPC).
MPC was originally used to handle constraints to achieve recursive feasibility\cite{mayneConstrainedModelPredictive2000}.
In some MPC applications, the prediction stages can forecast certain future \textit{external signal} values that impacts the system dynamics\cite{ZHANG2023111101}.
The word \textit{external} here means that this signal is neither a state nor an output of the system to be controlled.
In this paper, we call this external signal the \textit{generalized disturbance}.
For our problems of interest, we use the phrase predictive optimal control instead of MPC, mainly because we want to to emphasize that a future generalized disturbance has to be forecasted, and the major goal of the control is to minimize a certain cost.
Meanwhile, in this predictive optimal control framework, the prediction horizon and prediction update frequency is flexible.
This still fits the general MPC framework, but it may be different from the commonly-used receding horizon MPC.

In many predictive optimal control applications, the future to be predicted has intrinsic uncertainties\cite{vallon_data-driven_2022}. 
The to-be-predicted generalized disturbance may be future human maneuvers or ambient factors such as the temperature\cite{DEY2023100255}\cite{MUGNINI20232169}.
The future values of these signals are uncertain at the time of the forecast.
In practice, the prediction results may be in either stochastic or deterministic forms\cite{PANAGIOTELIS2023693}.
Many applications simply use deterministic predictions as it is easier to compute its corresponding optimal solution\cite{hu_multihorizon_2022}.
With more data and higher computing power, probabilistic forecast, also called stochastic prediction, is drawing more and more attentions from the control community.
Probabilistic forecast has been used in applications such as weather forecast\cite{palmerStochasticWeatherClimate2019}.
Different measures for evaluating stochastic predictions have been developed \cite{DensityForecastingSurvey}\cite{gneitingProbabilisticForecasting2014}.
There are techniques in Bayesian decision theory\cite{denoeuxDecisionmakingBeliefFunctions2019} \cite{petropoulosForecastingTheoryPractice2022} and stochastic MPC \cite{StochasticModelPredictive2016a}\cite{rosoliaDataDrivenPredictiveControl2018} utilizing a stochastic prediction for decision-making and control.

\begin{figure}[htbp]
\centerline{\includegraphics[width=\columnwidth]{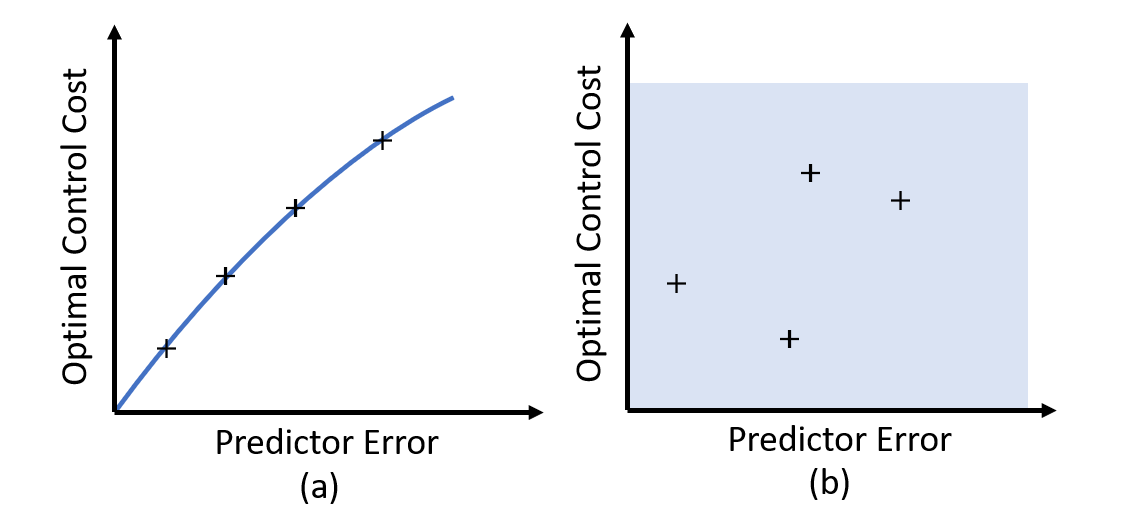}}
\caption{We sometimes assume a monotonic relationship like (a) between the predictor error and the optimal control cost. However, it is possible that the relationship is like (b), which means when the predictor improves, the control cost may get worse.}
\label{fig.cost_error}
\end{figure}

Since predictive optimal control has shown good performance, naturally, we are interested in investigating how we should design predictors and improve predictors. 
One important question is whether we can decouple the predictor design and the control system design.
In practice, we sometimes put a lot of efforts into improving the predictor, and hope this can lead to better predictive optimal control performance.
In this paper, the control performance is measured by the cost function defined in the optimal control.
When we try to improve the control by improving the prediction, the underlying assumption is that there is a monotonic relationship (like the one shown in Fig.\ref{fig.cost_error} (a)) between the predictor error and the optimal control cost.
However, in literatures, this assumption is rarely verified.
Actually, as we will show in this paper, the relation between the predictor and the optimal control performance is complicated.
Our theoretical analysis and numerical examples show that the relationship can be like Fig.\ref{fig.cost_error} (b), which means as the predictor improves, the optimal control cost may get worse.

The motivation of this paper is to provide tools to build connections between the predictor and the optimal control performance.
Prediction of a complicated process involving human behaviors or complex systems may never be perfect. 
We need to be able to analyze the predictive optimal control system while admitting that the prediction may be \textit{wrong}.
To thoroughly analyze this problem, we need to build an analysis framework describing the predictive optimal control process. 

Build a rigorous analysis framework for predictive optimal control is a nontrivial task for the following three reasons.
First, while it is easy to tell if a deterministic prediction is wrong, how we should handle the stochastic prediction case and define a \textit{truth} is not obvious, especially in scenarios where we cannot collect data repeatedly (i.e. it is almost impossible to re-create a driving scene with exactly the same environment, the same surrounding vehicles and drivers with the same status). 
Second, the ``prediction + optimal control'' scheme usually runs in a dynamic way with receding horizons, which means new prediction may override previous ones after an update, and new control policies may update accordingly.
This adds up to the complexity.
Third, the predictive optimal control involves a subjective predictor and an objective optimal control cost, which brings difficulties in notations. 
The complexities in describing the prediction performance and control performance are also somewhat related to the Bayesian vs. frequentist discrepancy in statistics.
The predictor usually relies on some \textit{a priori} assumptions, and the predicted probability distribution may be a subjective belief, which can be considered a Bayesian approach.
However, the control performance to be optimized is the objective long-run cost, which fits the frequentists' point of view.
In order to analyze predictive optimal control problems, we need to consider some perspectives from the Bayesians and frequentists simultaneously in one framework.

The tricky relationship between prediction performance and optimization performance has been noticed by some researchers, and it has been analyzed from data's point of views.
In ``machine-learning-based prediction + optimization'' problems, there is a growing interest in \textit{decision-focused learning}, which uses loss functions measuring the optimization results in the upstream machine learning prediction model training\cite{mandiSmartPredictandOptimizeHard2020}\cite{elmachtoubSmartPredictThen2020a}\cite{wilderMeldingDataDecisionsPipeline2019}.
For most dynamic system predictive optimal control applications, the predictor design is still separated from the optimal control process.
Meanwhile, there are many efforts on robust MPC focusing on generating good control policies despite imperfect predictions\cite{StochasticModelPredictive2016a}\cite{bayerRobustEconomicModel2016}.
However, formulations which can describe impact of wrong prediction for dynamic systems has not yet been reported in literatures.

In stochastic predictive optimal control, we need to deal with two types of descriptions of the future generalized disturbance: one predicted subjective probability distribution (which we call the \textit{belief}), and one ``true'' probability distribution (which will be precisely defined later).
In this paper, we focus on two questions:
\begin{itemize}
\item Q1 (What): What is the proper framework that describes the relationships among the predicted  probability distribution, the (to-be-defined) ``true'' probability distribution, the control performance, and other elements in predictive optimal control?
\item Q2 (How): With limited data and generally-unknown ``true'' probability distribution, how should we evaluate the predictors which generate the subjective probability distributions?
\end{itemize}
We believe Q1 has been answered in this paper, and a general answer to Q2 is provided.
More investigations of specific types of predictive optimal control will be needed for a comprehensive answer to Q2 in future studies. 

In this paper, we presents a framework describing the relationship among the elements of predictive optimal control that can be used to analyze the impact of wrong predictions.
We incorporate a stochastic environment model and define a new concept called the \textit{hidden prediction state} to connect the subjective belief and the objective truth.
Both the single-observation fixed-end-horizon case and the updating-observation receding-horizon case are considered.
Then we use this framework to consider the predictor evaluation problem, in the practical case with limited data availability.
Three commonly-used predictor evaluation measures, including the mean squared error, the regret and the log-likelihood, are discussed.
We show that a better predictor with respect to the mean squared error or the log-likelihood may actually lead to worse control performance, and this may happen even if the predictor is arbitrarily close to the global optimal.
Evaluating the predictor along with the control performance, such as using the control cost or the regret measure, can avoid this.
The results are illustrated in numerical examples and simulation examples from automotive applications.

The paper is structured as follows.
Section \ref{sec:problem} provides the general problem formulation of predictive optimal control.
Section \ref{sec:framework} presents an analysis framework for predictive optimal control with a environment model.
Section \ref{sec:evaluation} discusses predictor evaluation measures.
In Section \ref{sec.vs}, we use the proposed framework to analyze the relationship between the predictor performance and the control performance.
Section \ref{sec:examples} provides examples to illustrate the ideas and the results.
Section \ref{sec:conclusions} concludes this paper.

\section{The Predictive Optimal Control Problem}
\label{sec:problem}

\subsection{The Optimal Control Problem}

We consider the following discrete-time dynamic system
\begin{equation}\label{eq_sys_dyn}
x_{k+1} = f(x_k,w_k,u_k),
\end{equation}
where $x \in \mathbb{R}^{d_n}$, $w \in \mathbb{R}^{d_l}$,  and $u \in \mathbb{R}^{d_m}$. 
The state $x_k$ and the {\em generalized disturbance} $w_k$ can be directly measured at step $k$.
$w$ is usually considered as the output of a complex system, which is called the {\em environment}.
$u_k$ is the control input.

This dynamic system represents the physical system to be controlled, such as a vehicle, a robot, a machine, or a building.
The measured generalized disturbance $w$ may be the human input such as the pedal position and steering of the vehicle, or the ambient factors such as surrounding traffic behavior or the temperature.
We assume that $w$ has a relatively large impact on the dynamics and the cost.
Therefore, when we design the control $u$, we want to consider the impact of $w$.

We consider a finite-horizon problem of $N$ steps.
The cost $J$ in this finite horizon is the sum of running costs from step 0 to step $N-1$ and the terminal cost,
\begin{equation}
\begin{aligned}
	&J(x_0,w_0,u_0,...,x_{N-1},w_{N-1},u_{n-1},x_{N}) \\ 
	=&\sum_{k=0}^{N-1}l_k(x_k,w_k,u_k) + l_N(x_N).
\end{aligned}
\end{equation}
The goal is to find a {\em policy} $u = \pi(\cdot)$ to minimize the cost $J$.

If the complete {\em disturbance sequence} $\bar{w} = [w_0,w_1,...,w_{N-1}]$ (the brackets here mean an ordered sequence) is known at step 0, the problem can be solved as a deterministic optimal control problem. 
Many tools such as dynamic programming and Pontryagin’s minimum principle can be used to solve it.

However, in practice we usually do not know $\bar{w}$ in advance.
In our problem, $w_k$ is unknown before step $k$.
If we do not know the future values of $w$, we can no longer simply apply the deterministic optimal control.
Instead, we can consider this sequence $\bar{w}$ as a stochastic signal.
It can be represented by a random matrix of dimension $l \times N$, or equivalently, a random vector of dimension $lN$. 

The control objective is to find a feedback policy $u = \pi(\cdot)$ that uses available information to minimize the cost, or more rigorously, a certain {\em expectation} of the cost.
In the context of optimal control in this paper, \textit{better control performance} means a smaller cost expectation.

\subsection{Components of Predictive Optimal Control}

\begin{figure}[htbp]
\centerline{\includegraphics[width=\columnwidth]{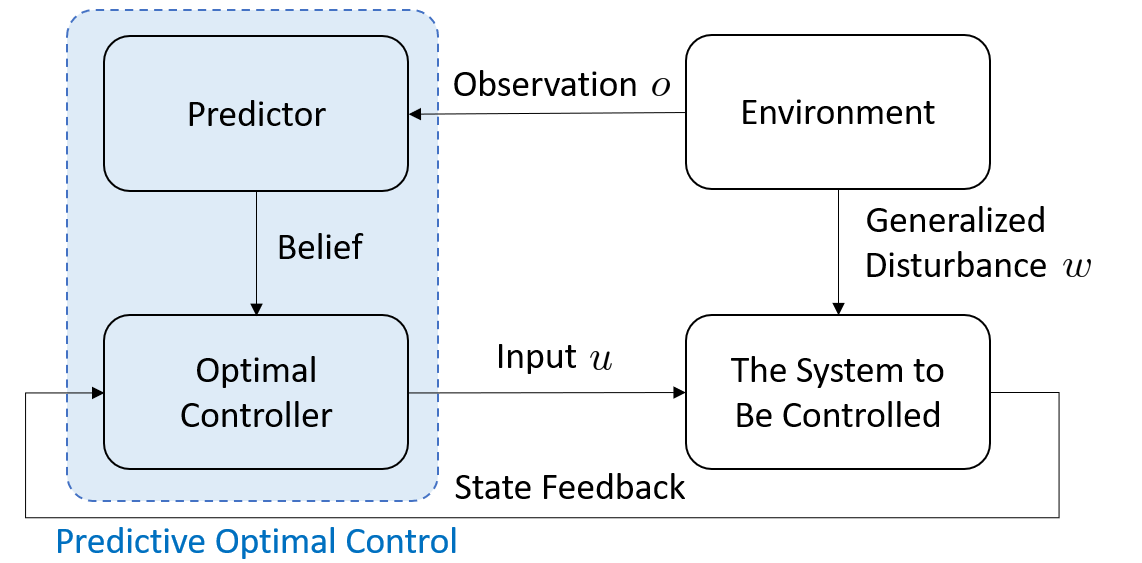}}
\caption{A typical predictive optimal control structure.
	A predictor generates a belief of the future based on observations of the environment. The optimal controller decides the control input based on the belief. }
\label{fig.poc_struc}
\end{figure}

To handle the future uncertainties of the disturbance sequence $\bar{w}$, it is common to use a predictor to forecast it.
We consider the general case where the forecast result is a discrete or continuous probability distribution of $\bar{w}$.
If one specific disturbance sequence is forecasted instead of a probability distribution, we may consider it as a distribution with a one or near-one probability at this specific sequence, and zero or near-zero probability at all other sequences. 
We call this probability distribution our \textit{belief}.

\begin{definition}[Belief]
A \textit{belief} is a subjective probability distribution of the future disturbance sequence $\bar{w}$.
\end{definition}

We use $\bar{W}_b$ to denote this probability distribution of $\bar{w}$.
$\bar{W}_b \in \bm{\bar{W}}$, where $\bm{\bar{W}}$ is the set of all possible probability distributions of $\bar{w}$.
We write $\bar{w} \sim \bar{W}_b$, which means $\bar{w}$ follows the distribution $\bar{W}_b$.
When there is no ambiguity, we do not distinguish the disturbance sequence probability distribution $\bar{W}_b$ and its data representation, which may be a high-dimensional vector to represent a probability mass function, or a vector of parameters for a probability density function.

To obtain a belief $\bar{W}_b$, we need to observe the environment for necessary information. 
We assume that $o \in \mathbb{O}$ is the observation from the environment.
The observation $o$ may be in the form of sensor readings, images, videos, or data received via communication, and their histories.
We can define the concept of \textit{predictors} as follows.

\begin{definition}[Predictor]
A \textit{predictor} $\mathcal{P}:\mathbb{O} \rightarrow \bm{\bar{W}}$ is a mapping from an observation to a belief. 
\end{definition}

With the belief generated by the predictor, we can compute the optimal control.
As we {\em believe} our {\em belief}, the expected cost to be minimized can be defined as 
\begin{equation}\label{eq.expected_cost}
\begin{aligned}
	& \mathbb{E} \, J(x_0,w_0,u_0,...,x_{N-1},w_{N-1},u_{n-1},x_{N}),
\end{aligned}
\end{equation}
where our belief tells us that $\bar{w} \sim \bar{W}_b$ and $\bar{W}_b = \mathcal{P}(o)$.
We use $\pi_{\bar{W}_b}(\cdot)$ to denote the optimal policy that minimizes the above expected cost when $\bar{w} \sim \bar{W}_b$.
This optimal control problem is well-defined, though computing the exact optimal policy $\pi_{\bar{W}_b}(\cdot)$ may be challenging if $\bar{W}_b$ is complicated.
In practice, an approximated solution is usually applied \cite{jingPredictiveOptimalControl2021}.

The predictive optimal control structure shown in Fig. \ref{fig.poc_struc} has been used in many control applications: first using a predictor to process the observation $o$ to obtain a belief $\bar{W}_b$, then computing  (or approximately computing)  the corresponding optimal control policy $u = \pi_{\bar{W}_b}(\cdot)$, finally applying the control to the dynamic system in (\ref{eq_sys_dyn}).
The conventional predictive optimal control formulation in practice just stops here, and it cannot provide further insights about the ``prediction + optimal control'' scheme. 
It assumes that $\bar{W}_b$ is \textit{correct}, so we use it in our optimal control design.
However, since our belief $\bar{W}_b$ is estimated, in most cases, it is actually different from the \textit{truth}.

\subsection{What If the Beliefs Are Wrong}

In real-world applications, usually we know that our belief is imperfect.
This brings up a lot of interesting questions.
In the rest of this paper, we will focus on the case where we know our belief is not perfect, and we would like to find out how it impacts the control performance, and how we can improve our it.

The \textit{true} probability distribution of the future disturbance sequence is denoted by $\bar{W}_t$.
We will define this true probability distribution in Section \ref{sec.truth}.
For now, let us assume there exists a well-defined one.

If we are not sure whether our \textit{belief} is \textit{true}, we have a problem immediately: since we do not know the true distribution $\bar{W}_t$, we cannot evaluate the cost expectation in (\ref{eq.expected_cost}) in the sense that $\bar{w} \sim \bar{W}_t$.
Based on our imperfect belief $\bar{W}_b$, we can only obtain an optimized policy with respect to this imperfect belief.
As $\bar{W}_b$ is only a subjective probability distribution which is not directly associated with any actual random vectors defined in our formulation so far, we use the notation $\mathop{\hat{\mathbb{E}}}_{\bar{w} \sim \bar{W}_b} *(\bar{w})$ to denote the \textit{believed expectation} of a function $*({\bar{w}})$ when the random vector $\bar{w}$ follows the \textit{believed distribution} $\bar{W}_b$.
We will stop using the ambiguous expectation notation in (\ref{eq.expected_cost}) for subject beliefs. 
Instead, the believed expected cost to be minimized is now defined as  
\begin{equation}\label{eq_believed_expected_cost}
\mathop{\hat{\mathbb{E}}}_{\bar{w} \sim \bar{W}_b} J(x_0,w_0,u_0,...,x_{N-1},w_{N-1},u_{n-1},x_{N}).
\end{equation}
Under this notation, the true expectation can be re-written as 
\begin{equation}\label{eq_true_expected_cost}
\mathop{\hat{\mathbb{E}}}_{\bar{w} \sim \bar{W}_t} J(x_0,w_0,u_0,...,x_{N-1},w_{N-1},u_{n-1},x_{N}).
\end{equation}

Our optimal policy $\pi_{\bar{W}_b}(\cdot)$ obtained using the imperfect belief $\bar{W}_b$ minimizes (\ref{eq_believed_expected_cost}), not (\ref{eq_true_expected_cost}).
It initially seems that there is not much we can do if our best estimation is $\bar{W}_b$.
However, a complete analysis framework will give us insights about the relationship between the predictor and the control performance, thus guide our predictor design.

\section{An Analysis Framework}
\label{sec:framework}

In this section, we present an analysis framework to consider the impact of imperfect predictions.
We will first introduce the environment model with the hidden predictions state, then discuss what a true probability distribution is, and finally integrate recurrent prediction schemes with the presented model.

\subsection{The Hidden Prediction State}

We assume that the environment that generates the disturbance $w$ is a dynamic system which can be described by the following equations,

\begin{equation}\label{eq_env}
\begin{aligned}
	z_{k+1} &= f_E(z_k,r_k) \\
	o_k &= h_{Eo}(z_k,r_k) \\
	w_k &= h_{Ew}(z_k,r_k),
\end{aligned}
\end{equation}
where $z_k$ is the state of the environment system, $r_0$, $r_1$, ..., $r_{N-1}$ are independent and identically distributed (i.i.d.) random disturbance, $o_k \in \mathbb{O}$ is the measured output, which is also called the observation of the environment, and $w_k$ is the output of the environment system that impacts our system to be controlled.
Since we assume that the value of $w_k$ is measured, $o_k$ contains all the information of $w_k$.
From the point of view of the system to be controlled, $w_k$ is the generalized disturbance.
Since the environment is usually very complex, such as the human beings and the weather system, the dimensions of $z$ and $r$ can be very high and the exact formulas of (\ref{eq_env}) may never be known.

\begin{figure}[htbp]
\centerline{\includegraphics[width=\columnwidth]{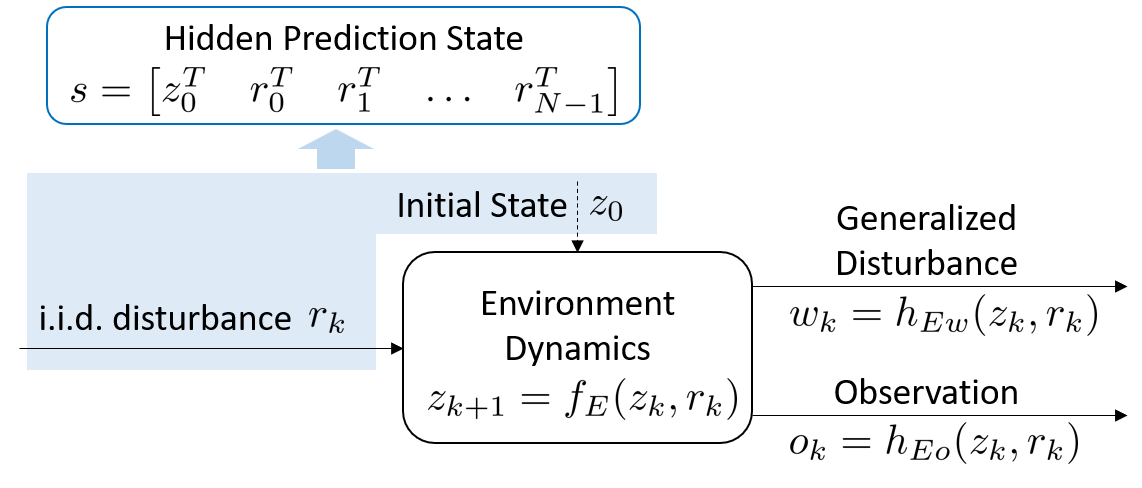}}
\caption{The hidden prediction state $s$ in the environment model. $s$ contains all factors from the environment that impacting the predictive optimal control process. It completely determines the realized disturbance sequence $\bar{w}$ and the observation sequence $\bar{o}$.}
\label{fig.env}
\end{figure}

Let us define the \textit{hidden prediction state}
\begin{equation}\label{eq_s}
s = \begin{bmatrix}
	z_0^T & r_0^T & r_1^T & \hdots & r_{N-1}^T
\end{bmatrix},
\end{equation} 
as shown in Fig. \ref{fig.env}.
According to the environment model in (\ref{eq_env}), the generalized disturbance sequence $\bar{w}$ can be completely determined by the hidden prediction state, which is a random vector, $s \sim S$.
Therefore, we write $\bar{w}(s)$ as $\bar{w}$ is uniquely determined when $s$ is given.
The observations $o$ can also be completely determined by $s$.
Therefore, we write
\begin{equation}
o_k(s) = h_{Eo}(z_k,r_k),
\end{equation}
and use the notations
\begin{equation}
\begin{aligned}
	o(s) &= o_0(s) = h_{Eo}(z_0,r_0) \\
	\bar{o}(s) &= [o_0(s),o_1(s),...,o_{N-1}(s)],
\end{aligned}
\end{equation}
for short.

In general, $s$ cannot be directly observed at step 0 due to two reasons.
First, the environment state $z_0$ and environment disturbance $r_0$ may not be estimated by only measuring $o_0$.
Second, the i.i.d. random disturbances $r_1$, $r_2$, ..., $r_{N-1}$ at future steps cannot be known at step 0.

Since there may be multiple ways building the environment model, the selection of the hidden prediction state $s$ is not unique.
In predictive optimal control applications, it is not necessary to formulate the environment and define the hidden prediction state at all.
However, the environment model in (\ref{eq_env}) and the definition of the hidden prediction state $s$ in (\ref{eq_s}) are the cores of the analysis framework, and they can help us in analyzing the problems when the predictions are not perfect.

\subsection{The True Probability Distribution}\label{sec.truth}

In the previous section, we assume that there is a true probability distribution $\bar{W}_t$.
Now let us define it with the environment model.

We consider the belief obtained at step 0.
With the environment in (\ref{eq_env}), our belief $\bar{W}_b$ is determined by applying the predictor $\mathcal{P}$ on the observation $o(s)$, therefore we can write 
\begin{equation}
\bar{W}_b = \mathcal{P}(o) = \mathcal{P}(o(s)).
\end{equation}
When there is no ambiguity, we may also write $\mathcal{P}(s)$ instead of $\mathcal{P}(o(s))$ for conciseness. 
We investigate multiple potential ways to define the true probability distributions.

\subsubsection{The A Posteriori Truth}
One may argue that we will be able to know the realized disturbance sequence $\bar{w}$ at step $N$, so the true probability of the realized disturbance sequence is one, and the probabilities of all other disturbance sequence are zero.
So for the belief $\bar{W}_b = \mathcal{P}(o(s))$, $\bar{w}(s)$ is the \textit{a posteriori} truth.

\subsubsection{The A Priori Truth}
One may argue that there are intrinsic uncertainties in $s$ as the future values of the i.i.d. environment disturbance $[r_1,r_2,...,r_{N-1}]$ can never be known at step 0. 
In a prediction, the true probability distribution should be determined by what has happened by the time of the prediction, not from the future. 
This probability distribution should be determined at step 0.
Therefore for the belief $\bar{W}_b = \mathcal{P}(o(s))$, the truth is the following conditional probability distribution
\begin{equation}
\bar{w}(S) \big | S_{z_0} = s_{z_0}, S_{r_0} = s_{r_0},
\end{equation}
where $s_{z_0}$, $s_{r_0}$ are the $z_0$ and $r_0$ component of $s$ respectively.
We call this \textit{a priori} true probability distribution $\bar{W}_{tn}(s_{z_0},s_{r_0})$, or $\bar{W}_{tn}(s)$ for short.

\subsubsection{The Observable Truth}
One may also argue that since our only observation of the environment is $o$, the best estimation of the distribution should be limited not just by the time of the prediction, but also by the information we have.
Therefore for the belief $\bar{W}_b = \mathcal{P}(o(s))$, the truth is the following conditional probability distribution
\begin{equation}
\bar{w}(S) \big | o(S) = o(s).
\end{equation}
We call this \textit{observable} (not related to the observability in control theory) true probability distribution $\bar{W}_{to}(o(s))$, or $\bar{W}_{to}(s)$ for short.

All the above three arguments about the true probability distributions make sense.
As $\bar{W}_{to}(o(s))$ is the \textit{observable} probability distribution that could be learned given enough data, we consider it as the benchmarking truth in this paper.
Making $\bar{W}_b = \mathcal{P}(o)$ close to $\bar{W}_{to}(o)$ is an intuitive way of improving the prediction, but how to define the difference between $\bar{W}_b$ and $\bar{W}_{to}(o)$ is an interesting question which will be discussed in Section \ref{sec:evaluation}.

\subsection{Recurrent Predictions}

In the predictive optimal control implementation under the structure in Fig. (\ref{fig.poc_struc}), the prediction runs recurrently at every control step.
Depending on whether new observations are used after the initial step and how long the prediction horizon is, we consider three typical types of recurrent predictions. 

\subsubsection{Type I: No Subsequent Observations After Step 0}

\begin{figure}[htbp]
\centerline{\includegraphics[width=\columnwidth]{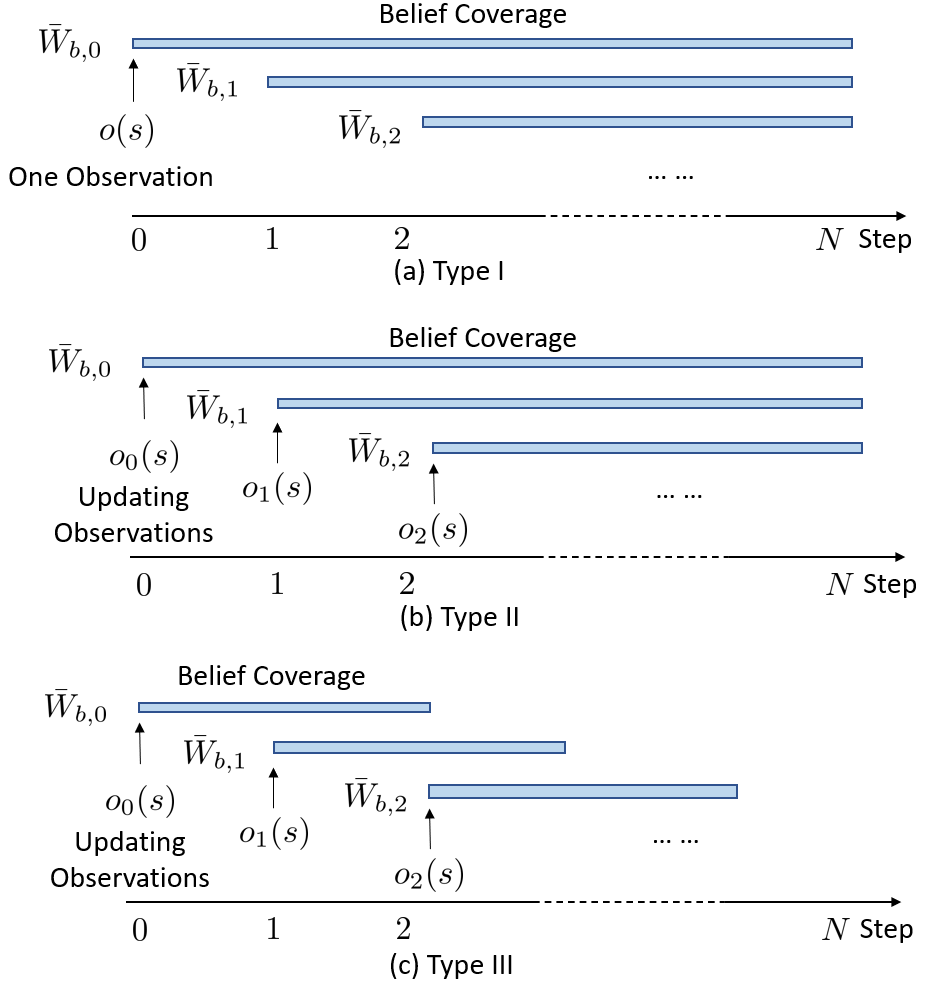}}
\caption{Three typical types of recurrent predictions. (a) Type I: the only observation is obtained at step 0. (b) Type II: a new observation is obtained at every step. (c). Type III: the prediction covers a fixed-length receding horizon with new observations at every step.}
\label{fig.types}
\end{figure}

\begin{figure}[htbp]
\centerline{\includegraphics[width=\columnwidth]{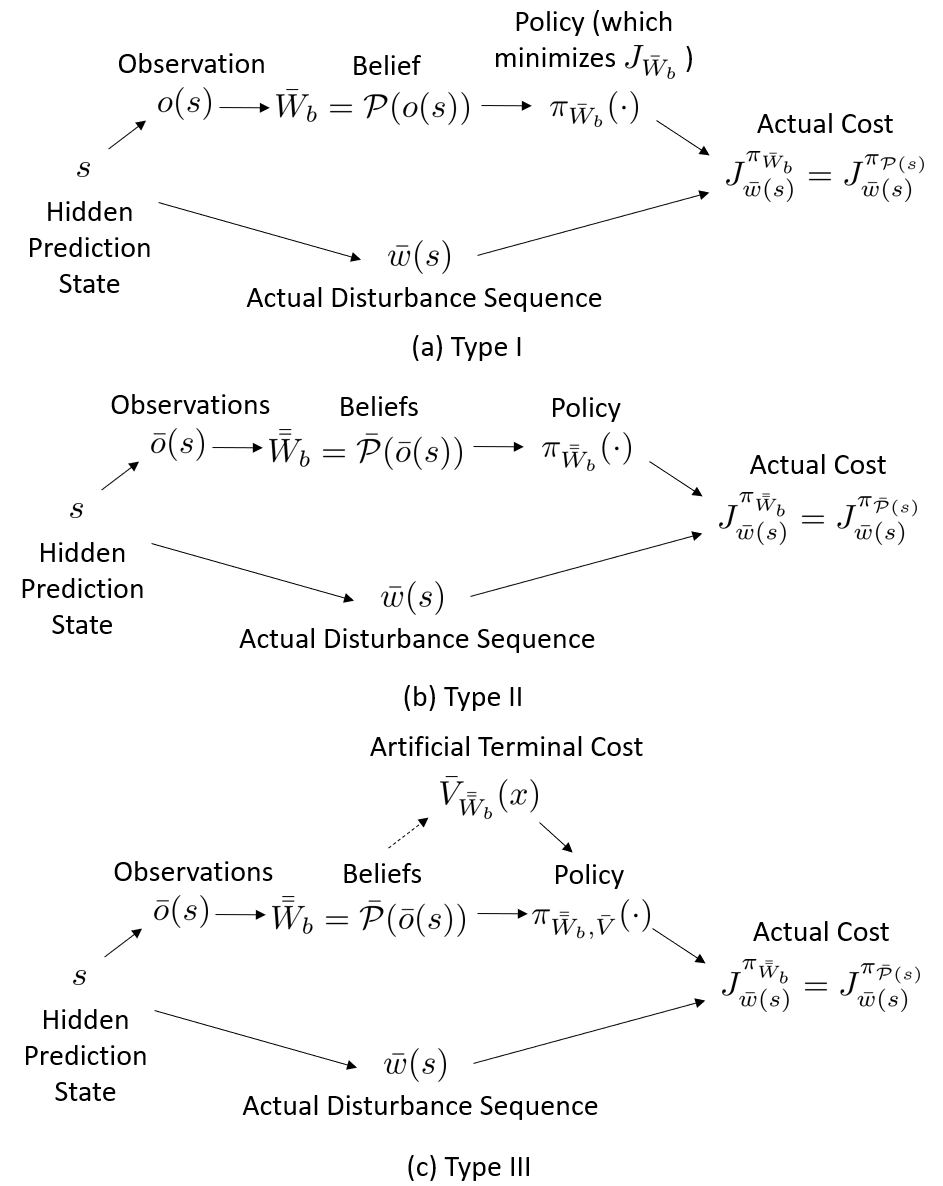}}
\caption{Impacters of the actual cost in predictive optimal control, when the predictor is given and the policy is optimal with respect to the belief. 
	(a) In Type I, the actual cost is uniquely determined by $s$ through two parallel paths: the controller path (the observation-belief-policy path) and the physics path (the disturbance path).
	(b) Paths in Type II are similar to Type I.
	(c) In Type III, the actual cost is determined by $s$ and the artificial terminal cost in the controller path.}
\label{fig.types_flow}
\end{figure}

We first consider the case where all information we obtain from the environment is $o_0$ and $w_0,w_1,...,w_N$, which means that there are no subsequent observations after step 0.
This type is commonly used when the process of obtaining the observation or forecasting is expensive, e.g., the observation is obtained in a pay-per-use way, or the initial forecasting computation takes a long time.
At step 0, given a predictor $\mathcal{P}$, our belief $\bar{W}_b = \mathcal{P}(o(s))$ is determined by the observation $o(s)$.
We denote this belief at step 0 as $\bar{W}_{b,0}$.
At step 1, even though no new observation of $o_1$ is available, $w_1$ is still measured so the forecasted belief can be updated according to the value of $w_1$.
This update is essentially computing a conditional probability given the original belief $\bar{W}_{b,0}$ and $w_1$.
In some context, this is also called a filtration process.
The filtration is repeated at every step. 
We denote the updated belief at step k as $\bar{W}_{b,k}$. The filtration update is 
\begin{equation}
\bar{W}_{b,k} = f_{W_I}(\bar{W}_{b,k-1},w_k).
\end{equation}
Each new belief contains the information of the probability distribution of the disturbance sequence from the current step to the terminal step, as shown in Fig. \ref{fig.types} (a).

Given a belief $\bar{W}_b$, the optimal policy $\pi_{\bar{W}_b}(\cdot)$ minimizes the expected cost defined in (\ref{eq_believed_expected_cost}).
The policy $\pi_{\bar{W}_b}(\cdot)$ can be obtained as a feedback control using all available information at each step.
All available information includes the history of the measured state $x$ and the measured generalized disturbance $w$.
Since (\ref{eq_sys_dyn}) tells us that when $w$ and $u$ are given, $x$ is Markovian. 
We can use the current $x_k$ in the feedback, instead of using all the history of $x$.
In general, the generalized disturbance $w$ is not Markovian.
Actually it can be considered as the output of a partially observable Markov decision process.
Therefore, we need to include the history of $w$ in the control feedback.
However, when the belief update is just a filtration, even if $w_k$ is not Markovian, $\bar{W}_{b,k}$ is Markovian.
Therefore, we can use $\bar{W}_{b,k}$ in the feedback, instead of the history of $w$ with the belief $\bar{W}_b$. 
The optimal policy $\pi_{\bar{W}_b}(\cdot)$ can be computed recursively using dynamic programming.
This feedback policy is in the form of
\begin{equation}
u_k = \pi_{\bar{W}_b}(k,x_k,[w_0,w_1,...,w_k]) = \pi_{\bar{W}_{b,k}}(k,x_k).
\end{equation}
$\pi_{\bar{W}_b}(\cdot)$ is completely determined by the initial belief ${\bar{W}_b}$, which means it is determined by the hidden prediction state $s$ along with the predictor $\mathcal{P}$.

We use $J^{\pi_{\bar{W}_b}}_{\bar{w}(s)}$ to denote the cost of system (\ref{eq_sys_dyn}) with a fixed initial state when the disturbance sequence is $\bar{w}(s)$ and the policy is the optimal policy $\pi_{\bar{W}_b}(\cdot)$ obtained with the belief $\bar{W}_b$ by minimizing (\ref{eq_believed_expected_cost}).
As shown in Fig. \ref{fig.types_flow} (a), given a predictor $\mathcal{P}$, this actual cost is completely determined by $s$.

We assume that any disturbance sequence with a non-zero probability in $\bar{W}_{tn}(s)$ has a non-zero probability in $\bar{W}_b$.
With this assumption, the policy $\pi_{\bar{W}_b}(\cdot)$ obtained using $\bar{W}_b$ can be applied when the true disturbance sequence distribution is $\bar{W}_{tn}(s)$.
This is why we may want to consider a deterministic prediction as a distribution with a near-one (instead of one) probability at this specific sequence, and near-zero (instead of zero) probability at all other sequences. 

\subsubsection{Type II: A New Observation at Every Step}

In Type II, a new observation is obtained at every step.
Therefore, the belief $\bar{W}_{b,k}$ is updated according to the new observation $o_k$ at every step.
Each belief is still about the disturbance sequence from the current step to the terminal step, as shown in Fig. \ref{fig.types} (b).
The belief is updated at every step using the new observation and the predictor,
\begin{equation}
\bar{W}_{b,k} = \mathcal{P}_k(o_k(s)),
\end{equation}
where the predictor $ \mathcal{P}$ may be dependent on $k$.
We denote the sequence of beliefs $[\bar{W}_{b,0},\bar{W}_{b,1},...,\bar{W}_{b,N-1}]$ as $\bar{\bar{W}}_b$, and we write 
\begin{equation}
\bar{\bar{W}}_b = \bar{\mathcal{P}}(\bar{o}(s)),
\end{equation}
as these beliefs are determined by the predictors and the sequence of observations $\bar{o}$.

In predictive optimal control, the control policy at step $k$ is determined using the belief $\bar{W}_{b,k}$, just like at any step of Type I.
So the at every step, the feedback control policy is in the form
\begin{equation}
u_k = \pi_{\bar{W}_{b,k}}(k,x_k).
\end{equation}
The policies in all steps form a sequence of policies $[\pi_{\bar{W}_{b,0}}(\cdot),\pi_{\bar{W}_{b,1}}(\cdot),...,\pi_{\bar{W}_{b,N-1}}(\cdot)]$.
Each policy $\pi_{\bar{W}_{b,k}}(\cdot)$ covers the time horizon from step $k$ to step $N-1$, with the prediction made at step $k$.
Each of them depends on their observation $o_k(s)$, thus ultimately depends on $s$.
At step $k$, we apply policy $\pi_{\bar{W}_{b,k}}(\cdot)$, then discard it since a new policy will be obtained at step $k+1$.
Combining the first steps (the step-$k$ where $k=0,1,...,N-1$) of each element of the policy sequence, we obtain one single policy, which can be denoted as $\pi_{[\bar{W}_{b,0},\bar{W}_{b,1},...,\bar{W}_{b,N}]}(\cdot)$.
We also write this policy as $\pi_{\bar{\bar{W}}_b}(\cdot)$, or equivalently, $\pi_{\bar{\mathcal{P}}(s)}(\cdot)$.
The the actual cost is $J^{\pi_{\bar{\bar{W}}_b}}_{\bar{w}(s)}$,  which can also be written as $J^{\pi_{\bar{\mathcal{P}}(s)}}_{\bar{w}(s)}$.
As shown in Fig. \ref{fig.types_flow} (b), this cost is determined uniquely by $s$ in a similar way as Type I.

\subsubsection{Type III: A New Observation at Every Step with Receding Horizon Prediction}

In Type III, the prediction is made on a time window of fixed-length in a receding horizon way.
This is commonly used in predictive optimal control applications. 
When the final step of the window is not $N$, an artificial terminal cost $V(\cdot)$ is usually designed for the optimal control problem.
The cost minimized at step $k$ is 
\begin{equation}
\begin{aligned}
	& \mathop{\hat{\mathbb{E}}}_{\bar{w} \sim \bar{W}_{b,k}} \big[ \sum_{i=k}^{k+n-1}l_k(x_i,w_i,u_i) + V_{\bar{W}_{b,k}}(x_{k+n}) \big],
\end{aligned}
\end{equation}
where $n$ is the window length.
The value of the artificial terminal cost $V_{\bar{W}_{b,k}}(x_{k+n})$ will impact the control policy, therefore it will impact the actual cost.
In this case, given the predictors, the actual cost is determined by both $s$ and the design of $V(\cdot)$.

We may consider the artificial terminal cost $V(\cdot)$ as an estimation of the optimal cost-to-go, that is, 
\begin{equation}
V_{\bar{W}_{b,k}}(x_{k+n}) 
\approx   \min \mathop{\hat{\mathbb{E}}}_{\bar{w} \sim \bar{W}_{b,k}}  \big[  \sum_{i=k+n}^{N-1}l_i(x_k,w_k,u_k) + l_N(x_N) \big].
\end{equation}
Then this type of problem can be considered as an approximation of Type II.
If this artificial terminal cost is exactly the optimal cost-to-go, it is the same as Type II.

\section{Predictor Evaluation}
\label{sec:evaluation}

In this section, we use the developed framework to discuss predictor evaluation in the context of predictive optimal control.
We will focus on the Type I problem, which is the foundation of all predictive control problems.

\subsection{Just Being Accurate Is Not Enough}

An accurate predictor means that when the predictor says the probability distribution of $\bar{w}$ is $\bar{W}_b$, the distribution is indeed $\bar{W}_b$.
However, just being accurate is not good enough for predictors.
To show this, we will define it rigorously first.

\begin{definition}[Maximum Indistinguishable Observation Set]
Give a predictor $\mathcal{P}$ and a belief $\bar{W}_{b}$, the \textit{maximum indistinguishable observation set} $O_M$, is the set of all observations based on which the prediction output is $\bar{W}_b$,
\begin{equation}
	O_M = \{o | \mathcal{P}(o) = \bar{W}_{b}\}.
\end{equation}

\end{definition}

\begin{definition}[Accurate]
A predictor $\mathcal{P}$ is \textit{accurate} over an observation set $\mathbb{O}$ if belief $\bar{W}_{b}=\mathcal{P}(o_i)$ generated by this predictor over any observation $o_i \in \mathbb{O}$, equals the probability distribution of $\bar{w}(S)$ conditioned on $o(S) \in O_M$, where $O_M$ is the maximum indistinguishable observation set of $\mathcal{P}$ and $\bar{W}_{b}$.
\end{definition}

This definition means that if we collect all realized disturbance sequence data for any fixed belief from an accurate predictor, the collected data distribution will match this belief.
Simply being accurate does not mean a good predictor.
For example, a \textit{blind} predictor is a predictor that generates the same belief for all observations.
An \textit{accurate} blind predictor does not use any information from the observation $o$, yet it offers accurate beliefs.
It is accurate, but not very informative.

\subsection{The Goal is $\bar{W}_{to}(o)$, But Getting There Is A Challenge}

The goal of prediction is to obtain the true distribution $\bar{W}_{to}(o)$ for every \textit{observation} $o$.
To show this, we will prove that if our belief is the same as $\bar{W}_{to}(o)$, we obtain the optimal control performance, as long as we use no more information about $s$ than $o(s)$.

Given a predictor $\mathcal{P}$, the predictive optimal control cost expectation is $\mathbb{E} \, J^{\pi_{\mathcal{P}(S)}}_{\bar{w}(S)}$.
By the law of total expectation and the definition of $W_{to}(o(s))$, 
\begin{equation}
\mathbb{E} \, J^{\pi_{\mathcal{P}(S)}}_{\bar{w}(S)} = 
\mathbb{E} \, \Big[\mathbb{E}[J^{\pi_{\mathcal{P}(S)}}_{\bar{w}(S)} | o(S)] \Big]=
\mathbb{E} \,\Big[  \mathop{\hat{\mathbb{E}}}_{\bar{w} \sim \bar{W}_{to}(o(S))} J^{\pi_{\mathcal{P}(S)}}_{\bar{w}} \Big].
\end{equation}
By definition, $\pi_{\bar{W}_{to}(o)}$ is the optimal strategy that minimizes $\mathop{\hat{\mathbb{E}}}_{\bar{w} \sim \bar{W}_{to}(o)} J^{(\cdot)}_{\bar{w}}$
for every $o$.
Since we use no more information about $s$ than $o(s)$, given any $o$, this cost is the best that we can achieve.
Therefore, given any $o$, $\bar{W}_{to}(o)$ minimizes the predictive optimal control actual cost $\mathop{\hat{\mathbb{E}}}_{\bar{w} \sim \bar{W}_{to}(o)} J^{\pi_{(\cdot)}}_{\bar{w}}$.
So it minimizes the predictive optimal control cost expectation.

The goal of our predictor result $\mathcal{P}(o)$ is $\bar{W}_{to}(o)$.
However, the path to achieve this goal is full of challenges.
First, in practice, we may never be able to directly compare the predicted belief $\mathcal{P}(o)$ with our target $\bar{W}_{to}(o)$ as $\bar{W}_{to}(o)$ is unknown.
Second, in general, there is no guarantee that predictors generating beliefs closers to $\bar{W}_{to}(o)$ lead to better control performance.
Even locally around $W_{to}(o)$, as long as $\mathcal{P}(o) \neq \bar{W}_{to}(o)$, we do not have such guarantees (details will be discussed in Section \ref{sec.vs} and examples are provided in Section \ref{sec:examples}).
Nevertheless, this ultimate goal still provides us an incentive to keep optimizing the predictor towards $\bar{W}_{to}(o)$.

\subsection{Available Data for Evaluation}

In most predictive optimal control applications, it is not possible to obtain enough data to estimate $\bar{W}_{to}(o)$ for every $o$.
In practice, the data samples are in the form of $\{o,\bar{w}\}$ pairs.
For a few specific $o^i$, we may have many $\{o^i,\bar{w}\}$ pair samples such that $\bar{W}_{to}(o^i)$ can be directly estimated.
However, for most $o^i$, we usually do not have enough $\{o^i,\bar{w}\}$ samples to estimate $\bar{W}_{to}(o^i)$.
For example, in human-driven vehicle speed prediction, where the to-be-predicted $w$ is the vehicle speed, and the observation $o$ is the driving scenario data including the vehicle status, driver status, road conditions, and traffic conditions, etc.
We can collect many samples in the form of $\{o,\bar{w}\}$ pairs under many driving scenarios.
But for a specific driving scenario, it is very difficult to have repeated data showing how a driver behaves differently each time, as the scenario keeps changing.
In practical situations, we usually make some assumptions on the predictor and parameterize the predictor so that it can be trained using $\{o,\bar{w}\}$ pairs.
We also need to evaluate predictor performance using many $\{o,\bar{w}\}$ pairs, while not assuming we have repeated data for every specific $o^i$ to learn $\bar{W}_{to}(o^i)$.

\subsection{Predictor Evaluation Using Available Data}

With one $\{o,\bar{w}\}$ pair, given a predictor $\mathcal{P}$, we can compute a belief $\bar{W}_b = \mathcal{P}(o)$.
Using a beleif $\bar{W}_b$ and the corresponding realized disturbance sequence $\bar{w}$, we can define some one-time prediction performance measures.
With multiple $\{o,\bar{w}\}$ pairs, the predictor $\mathcal{P}$'s performance can be evaluated by aggregating these one-time prediction performance measures.

\subsubsection{One-Time Prediction Performance Measures}
We consider two types of measures: the error-based measures, and the probability-based measures.
Given a realized disturbance sequence $\bar{w} \in\bm{\bar{w}}$ and a belief $\bar{W}_b \in \bm{\bar{W}}$, the one-time prediction performance measure is in the form of $m:\bm{\bar{w}} \times \bm{\bar{W}}\rightarrow \mathbb{R}$.
If $m$ satisfies that 
\begin{equation}
m(\bar{w},\bar{W}_b) \geq 0, \text{ for all } \bar{w} \text{ and } \bar{W}_b,
\end{equation}
and 
\begin{equation}
m(\bar{w},\bar{W}_b) = 0 \text{ if } \hat{\mathbb{P}}_{\bar{W}_b}(\bar{w})=1,
\end{equation}
where $\hat{\mathbb{P}}_{\bar{W}_b}(\bar{w})$ means the probability of $\bar{w}$ in the belief $\bar{W}_b$,
we say $m$ is an error-based measure.
Furthermore, when $m(\bar{w},\bar{W}_b) = 0$ only if $\hat{\mathbb{P}}_{\bar{W}_b}(\bar{w})=1$, this error measure is called a \textit{strict} one-time prediction error measure.
With a slight abuse of notation, we write $\bar{W}_b = \bar{w}$ if $\hat{\mathbb{P}}_{\bar{W}_b}(\bar{w})=1$.
Then $\bar{W}_b = \bar{w}$ minimizes any error-based measure $m(\bar{w},\cdot)$ by definition.

Here are two examples of error-based measures: the expected mean squared error (MSE), and the regret.
The commonly-used MSE, 
\begin{equation}
M(\bar{w},\bar{W}_b) = \mathop{\hat{\mathbb{E}}}_{\bar{w}_b \sim \bar{W}_b}
\norm{\bar{w} - \bar{w}_b}_2^2,
\end{equation} 
is a strict one-time prediction error measure.
The expectation here is the subjective expectation with respect to the belief $\bar{W}_b$.
The regret
\begin{equation}
R(\bar{w},\bar{W}_b) = J^{\pi_{\bar{W}_b}}_{\bar{w}} - J^{\pi_{\bar{w}}}_{\bar{w}},
\end{equation}
is a non-strict one-time prediction error measure. 
It is defined as the difference between the optimal cost with the predicted belief, and the optimal cost with the \textit{a posteriori} disturbance sequence.
Computing the regret is usually much more difficult than computing the MSE, as it involves solving a stochastic optimal control problem.
It is essentially evaluating the control performance of a specific dynamic system with this prediction, instead of evaluating the prediction alone.

Besides the error measures, we may also use a probability-based measure such as the log-likelihood,
\begin{equation}
P(\bar{w},\bar{W}_b) = \begin{cases}
	-\log \hat{\mathbb{P}}_{\bar{W}_b}(\bar{w}), \, &\text{if $\bar{W}_b$ is discrete}, \\
	-\log \hat{\mathbb{F}}_{\bar{W}_b}(\bar{w}), \, &\text{if $\bar{W}_b$ is continuous},
\end{cases} \\
\end{equation}
where $\hat{\mathbb{P}}_{\bar{W}_b}(\bar{w})$ is believed probability of the realized disturbance sequence, and $\hat{\mathbb{F}}_{\bar{W}_b}(\bar{w})$ is the believed probability density at the realized disturbance sequence.
We use the log-likelihood because we will sum up or average these one-time prediction performance to evaluate the predictor performance.
There is a minus sign because we want to keep this probability-based measure consistent with the error-based measures, which are to be minimized.

\subsubsection{Predictor Performance Measures}
If we compute the average of multiple one-time prediction performance measures from one predictor's result, we obtain a predictor performance measure, whose expectation is 
\begin{equation}\label{eq.exp_PE}
\begin{aligned}
	\mathcal{E}_m(\mathcal{P})=& \mathbb{E} \, m[\bar{w}(S),\mathcal{P}(o(S))] \\
	=&  \mathbb{E} \, \Big[\mathbb{E} \, \Big[m[\bar{w}(S),\mathcal{P}(o(S))] \Big | o(S)\Big]\Big].
\end{aligned}
\end{equation}
Based on our definition of the \textit{observable} true distribution $\bar{W}_{to}(o)$, for a given, fixed observation $o$, the expectation of the one-time prediction performance measure of a belief $\bar{W}_b$ can be computed using $\bar{W}_{to}(o)$, 
\begin{equation}\label{eq.exp_EE}
\begin{aligned}
	\mathbb{E} \, \Big[m[\bar{w}(S),\bar{W}_b] \Big | o(S) = o \Big] &= \mathop{\hat{\mathbb{E}}}_{\bar{w} \sim \bar{W}_{to}(o)}  m(\bar{w},\bar{W}_b).
\end{aligned}
\end{equation}
The right-hand side of the (\ref{eq.exp_EE}) can be computed just using two probability distributions $\bar{W}_{to}(o)$ and $\bar{W}_b$.
Therefore, we define 
\begin{equation}\label{eq.E_e}
\begin{aligned}
	E_m(\bar{W}_{to}(o),\bar{W}_b)= \mathop{\hat{\mathbb{E}}}_{\bar{w} \sim \bar{W}_{to}(o)}  m(\bar{w},\bar{W}_b).
\end{aligned}
\end{equation}

According to (\ref{eq.exp_PE}), (\ref{eq.exp_EE}) and (\ref{eq.E_e}), given a one-time prediction performance measure $m$, the expected predictor performance measure can also be written as 
\begin{equation}\label{eq.EmE}
\begin{aligned}
	\mathcal{E}_m(\mathcal{P}) =&  \mathbb{E} \,  E_m[\bar{W}_{to}(o(S)),\mathcal{P}(o(S))] .
\end{aligned}
\end{equation}

\section{Predictor Performance vs. Control Performance}\label{sec.vs}

We are interested in the following two properties of predictor measures: 
(1) \textit{best-P-lowest-C}: whether the best performed predictor always leads to the lowest predictive optimal control cost,
(2) \textit{better-P-lower-C}: whether better performed predictors always lead to lower predictive optimal control costs.
The performance and cost here mean the expectation of the performance measure and the cost. 
Since the predictor performance measures are in the form of (\ref{eq.EmE}), we analyze this by investigating different predictor performance measures under each given observation $o$.

\subsection{MSE: A Poor Measure}
if we choose the expected MSE as the one-time prediction performance measure, then, 
\begin{equation}
\begin{aligned}
	E_M(\bar{W}_{to},\bar{W}_b)
	=& \mathop{\hat{\mathbb{E}}}_{\bar{w} \sim \bar{W}_{to}}  M(\bar{w},\bar{W}_b)\\ 
	=& \mathop{\hat{\mathbb{E}}}_{\bar{w} \sim \bar{W}_{to}}  \Big[ \mathop{\hat{\mathbb{E}}}_{\bar{w}_b \sim \bar{W}_b}
	\norm{\bar{w} - \bar{w}_b}_2^2 \Big].
\end{aligned}
\end{equation}
In this case, $\bar{W}_b = \bar{W}_{to}$ may not even locally minimize $E_M(\bar{W}_{to},\cdot)$.
This means that when using MSE, even with a large amount of data, we will miss the preditor's ultimate goal $\bar{W}_{to}$.
When there is no constraint, the best $\bar{W}_b$ may be a deterministic prediction, which is inferior to the best stochastic predictor in terms of the control performance.
With the proposed framework, examples can be easily constructed to show that the MSE-based measure is neither a best-P-lowest-C measure nor a better-P-lower-C measure (see Section \ref{sec:examples}).

\subsection{Regret: A Good But Computationally Expensive Measure}
If the regret-based predictor performance measure is used, then 
\begin{equation}
\begin{aligned}
	E_R(\bar{W}_{to}(o),\bar{W}_b) 
	=& \mathop{\hat{\mathbb{E}}}_{\bar{w} \sim \bar{W}_{to}(o)}  R(\bar{w},\bar{W}_b)\\ 
	=& \mathop{\hat{\mathbb{E}}}_{\bar{w} \sim \bar{W}_{to}(o)}  
	(J^{\pi_{\bar{W}_b}}_{\bar{w}} - J^{\pi_{\bar{w}}}_{\bar{w}}) \\ 
	=& \mathop{\hat{\mathbb{E}}}_{\bar{w} \sim \bar{W}_{to}(o)}  
	J^{\pi_{\bar{W}_b}}_{\bar{w}} - \mathop{\hat{\mathbb{E}}}_{\bar{w} \sim \bar{W}_{to}(o)}  J^{\pi_{\bar{w}}}_{\bar{w}} .
\end{aligned}
\end{equation}
The second term $\mathop{\hat{\mathbb{E}}}_{\bar{w} \sim \bar{W}_{to}}  J^{\pi_{\bar{w}}}_{\bar{w}}$ is the \textit{posteriori} optimal cost, which is a constant given $\bar{W}_{to}$.
The first term is the same as the expected cost.
The regret is essentially the control cost with a constant offset.
$\bar{W}_{to}$ globally minimizes $E_R(\bar{W}_{to}(o),\cdot)$.
Furthermore, decreasing $E_R(\bar{W}_{to}(o),\bar{W}_b)$ will lead to a decrease of the expected cost.
Evaluating the predictors using regret-based error measures is essentially evaluating the control performance after connecting the prediction and the optimal control process.
Actually, we can just compute the first term $\mathop{\hat{\mathbb{E}}}_{\bar{w} \sim \bar{W}_{to}(o)}  
J^{\pi_{\bar{W}_b}}_{\bar{w}}$, which is the control cost under the predictor, to evaluate this predictor.
The regret is both a best-P-lowest-C and a better-P-lower-C measure.

\subsection{Log-Likelihood: A Probably-Fine Measure}
If we use the probability-based measure log-likelihood $P$, in the discrete case,
\begin{equation}
\begin{aligned}
	E_P(\bar{W}_{to},\bar{W}_b)
	=& \mathop{\hat{\mathbb{E}}}_{\bar{w} \sim \bar{W}_{to}}  P(\bar{w},\bar{W}_b) \\
	=& - \mathop{\hat{\mathbb{E}}}_{\bar{w} \sim \bar{W}_{to}} \log \hat{\mathbb{P}}_{\bar{W}_b}(\bar{w}) \\
	=& - \sum_{\bar{w}} \mathbb{P}(\bar{w}|o) \log \hat{\mathbb{P}}_{\bar{W}_b}(\bar{w}).
\end{aligned}
\end{equation}
This expectation is related to two probability distributions: the subjective belief and the \textit{observable} true probability distribution 
(it is different from the entropy $-\sum_{\bar{w}} \hat{\mathbb{P}}_{\bar{W}_b}(\bar{w}) \log \hat{\mathbb{P}}_{\bar{W}_b}(\bar{w})$, which describes the property of one probability distribution).
The best predictor is $\hat{\mathbb{P}}_{\bar{W}_b}(\bar{w}) = \hat{\mathbb{P}}_{\bar{W}_{to}}(\bar{w})$ for all $\bar{w}$ due to convexity, which means this measure is best-P-lowest-C.
However, there is no guarantee to make it better-P-lower-C.

\subsection{Summaries}
A summary of the three predictor measures is provided in Table \ref{tab1}.
Neither the MSE nor the log-likelihood measure have the Better-P-lower-C property.
For a general predictive control problem, , there is no guarantee that predictors with better MSE or log-likelihood lead to better control performance.
It implies that the predictor design cannot be simply decoupled from the downstream optimal control problem, and the predictor needs to be evaluated along with the control system performance, e.g., using the control cost or the regret as the predictor performance measure.

\begin{table}
\caption{Comparing Predictor Measures}
\label{table}
\setlength{\tabcolsep}{7pt}
	\begin{tabular}{|c|c|c|c|}
		\hline
		Predictor Measure& 
		Best-P-lowest-C& 
		Better-P-lower-C&
		Computation\\
		\hline
		MSE& No& No& Low \\
		Regret& Yes& Yes& High \\
		Log-likelihood& Yes& No& Low \\
		\hline
	\end{tabular}
\label{tab1}
\end{table}


\section{Illustrative Examples}
\label{sec:examples}

\subsection{A Simple Linear System Example}

We provide a numerical example where the predictor measure gets better while the control cost gets worse in a Type I problem, and illustrate the differences of the three predictor measures shown in Table \ref{tab1}.
Consider the simple linear system,
\begin{equation}
x_{k+1} = x_k + w_k + u_k, \quad k=0,1,
\end{equation}
where $-1 \leq u \leq 1$, $x_0 = w_0 = 0$,
with a quadratic cost function $J = x_2^2$.
There is no observation information other than $w$.
The input $u_1$ will be determined after $w_1$ is measured.
So the key of this predictive optimal control is to forecast the value of $w_1$ and determine $u_0$ accordingly. 

Assume that $w_1 \in \{-3,2\}$.
Therefore, the \textit{observable} true probability distribution of $w_1$ is in the form of
\begin{equation}
\mathbb{P}(w_1 = -3) = p, \quad \mathbb{P}(w_1 = 2) = 1-p.
\end{equation}
In addition, we assume that $0 \leq p \leq \frac{2}{3}$.
In this example, the hidden prediction state $s$ can be considered as $w_1$.
If we know the value of $p$ and can use it to design the optimal control accordingly, the ideal optimal policy is
\begin{equation}\label{eq.eg_sol}
\begin{aligned}
	u_1^* &= 
	\begin{cases}
		1, \quad &\text{if }  w_1 = -3, \\
		-1, \quad &\text{if }  w_1 = 2,
	\end{cases} \\
	u_0^* &= 3p-1,
\end{aligned}
\end{equation}
and the ideal optimal cost expectation is,
\begin{equation}
\mathbb{E} \, J^{\pi_{\bar{W}_{to}(o(S))}}_{\bar{w}(S)} = -9p^2+9p.
\end{equation}

A predictor gives a prediction of $p$ as $p_b$ and $0 \leq p_b \leq \frac{2}{3}$.
$p_b$ generates a belief of $w_1$.
The predictive optimal solution is in the same form as (\ref{eq.eg_sol}) while replacing $p$ with $p_b$.
With this belief, the predictive optimal control cost expectation is
\begin{equation}
\mathbb{E} \, J^{\pi_{\mathcal{P}(o(S))}}_{\bar{w}(S)} = 9p_b^2 - 18pp_b + 9p.
\end{equation}
Obviously, $p_b = p$ minimizes this cost expectation.

When using the MSE, the regret, and the log-likelihood measure, the predictor performance measures are
\begin{align}
	E_M(\bar{W}_{to},\bar{w}_b) &= -50pp_b + 25p_b + 25p, \\
	E_R(\bar{W}_{to},\bar{w}_b) &= 9p_b^2 - 18pp_b + 8p, \\
	E_P(\bar{W}_{to},\bar{w}_b) &= -p\log p_b-(1-p) \log (1-p_b),
\end{align}
respectively.
In the regret and the log-likelihood, $p_b= p$ minimizes the measures, as they are best-P-lowest-C measures.
Furthermore, for the regret, $E_R(\bar{W}_{to},\bar{w}_b)$ is just the cost expectation $\mathbb{E} \, J^{\pi_{\mathcal{P}(o(S))}}_{\bar{w}(S)}$ with a constant offset. 
However, in MSE, depending on the sign of $1-2p$, $p_b$ takes its minimal MSE value at $0$ or $\frac{2}{3}$, which is different from its optimal result $p$ in terms of the optimal control cost.

\begin{figure}[htbp]
\centerline{\includegraphics[width=\columnwidth]{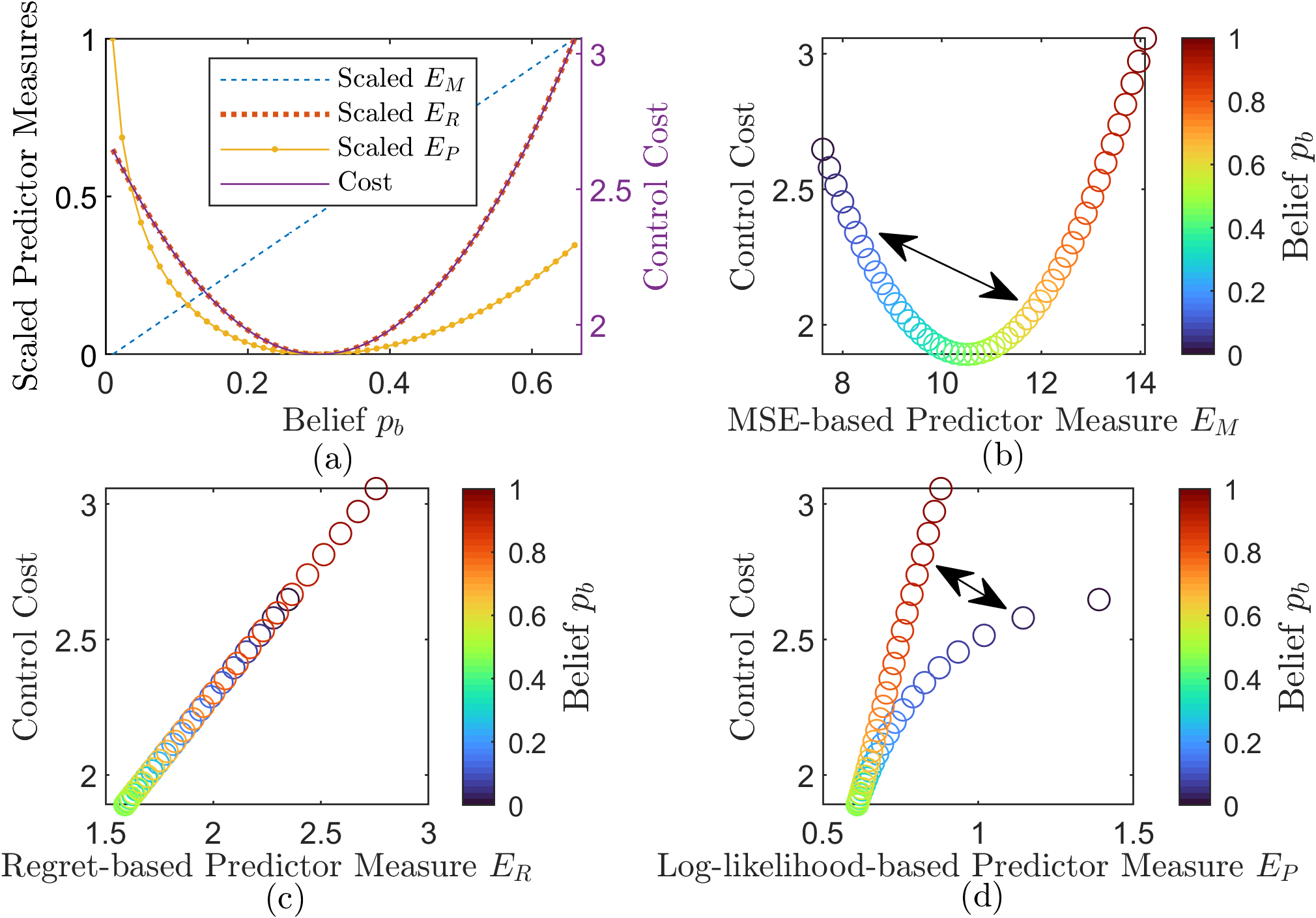}}
\caption{The change of the cost and the predictor measures as the belief changes when $p=0.3$. 
	In (a), all the predictor measures are scaled to $[0,1]$. 
	The cost and the scaled $E_R$ are overlapped as they have exactly the same shape. 
	(b) (c) and (d) each shows the relationship between a predictor measure and the control cost. In (b) and (d), the arrows indicate cases where the predictor measure improves while the cost gets worse. 
	Such pair exists even if the predictor is arbitrarily close to the global optimal.}
\label{fig.eg}
\end{figure}

We use $p=0.3$ to illustrates the better-P-lower-C properties of the measures.
The trend of the predictor measures and the cost expectation is shown in Fig. \ref{fig.eg} as the belief $p_b$ changes.
In the MSE and the log-likelihood case, it is possible to improve the predictor measure while making the optimal control cost worse.

\subsection{Automotive Examples with Real-World Driving Data}

We use a hybrid electric vehicle example in the form of a Type III problem to demonstrate the relationship between predictors and control performance in the real world.
Given a pre-defined driving cycle as a time-velocity table, optimal control can be used to determine the most efficient energy management strategy for a hybrid electric vehicle.
With uncertain future driving cycles, predictive optimal control can be used and the future vehicle velocities are forecasted.
We use real-world driving data from the Next Generation Simulation (NGSIM) \cite{TransportationFederalHighwayAdministration2016}.
We consider the energy management strategy for a simple hybrid electric vehicle powertrain shown in \cite{zengOptimizingEnergyManagement2017}.
The hybrid electric vehicle powertrain model is as follows, 
\begin{equation}
\begin{aligned}
	x_{SOC,k+1}=&x_{SOC,k} -K\eta_b \eta_m \omega_k T_{m,k},
\end{aligned}
\label{Dynamic function_soc}
\end{equation}
where $\eta_b = \eta_b(x_{SOC,k},\omega_k,T_{m,k})$, $\eta_m= \eta_m(\omega_k,T_{m,k})$, $x_{SOC}$ is the battery state of charge (SOC), $K$ is a constant, $\eta_b$ is a variable representing the battery or the inverse of the battery efficiency, depending on the sign of the motor power, $\eta_m$ represents the motor efficiency or the inverse of the motor efficiency, $\omega$ is the motor speed, $T_m$ is the motor torque.
We assume the following static relationship is known,
\begin{equation}
\begin{aligned}
	[\omega, T_{d}]^T= f_s(v,a), 
\end{aligned}
\end{equation}
where $T_d$ is the total torque demand to the powertrain, $v$ is the vehicle velocity, $a$ is the vehicle acceleration.
The control input is the motor torque $T_m$.
Once $T_m$ is determined, the engine torque $T_e$ is determined as 
\begin{equation}
\begin{aligned}
	T_e = T_d - T_m.
\end{aligned}
\end{equation}
The cost function is defined as 
\begin{equation}
\begin{aligned}
	J =  & \alpha_1 (x_{soc,N} - x_{soc,0}) \\ 
	& + \sum_{k=0}^N [\alpha_2 J_{FC,k}(\omega_k,T_{e,k}) + \alpha_3 (x_{soc,k}-x_{target})^2],
\end{aligned}
\end{equation}
where $\alpha_1$, $\alpha_2$ and $\alpha_3$ are weights, $J_{FC}$ is the fuel consumption determined by $\omega$ and $T_e$, and $x_{target}$ is the target battery SOC to be maintained.
This cost function considers the electric energy consumed, the fuel consumed, and the battery SOC variation from the target value during the $N$ steps.

In this problem, $w = [v,a]^T$ (or equivalently, $w = [\omega,T_d]^T$) is the generalized disturbance. 
Observation $o$ is just $w$ and its history.
A predictor tries to forecast future $w$, that is, the future velocity and acceleration. 
The future velocity and acceleration is dependent on various factors including the driving style, the road and traffic condition. 
In the proposed analysis framework, these factors are considered in the environment dynamics equations in (\ref{eq_env}).
The hidden prediction state $s$ associated with this environment dynamic system may be a high-dimensional vector.
But the exact formulas associated with the environment and the hidden prediction state are not needed for applying predictive optimal control.


Four types of deterministic predictors and two types of stochastic predictors are designed.

(D1) Constant velocity predictor.
\begin{equation}
a_{k+1} = 0.
\end{equation}

(D2) Linear decay acceleration predictor.
\begin{equation}
a_{k+1} = a_{k}-\frac{a_0}{\gamma}, \\
\label{Deterministic acceleration attenuation predictor}
\end{equation} 
where $\gamma$ is a constant parameter.

(D3) Exponential decay acceleration predictor.
\begin{equation}
a_{k+1} = \lambda a_k,
\end{equation}
where $\lambda \in (0,1)$ is a constant parameter.

(D4) Deterministic Long Short-Term Memory (LSTM) predictor,
\begin{equation}
\bar{v}^+ = f_{LSTM}(\bar{v}^-),
\end{equation}
where $\bar{v}^+ = [v_{k+1},v_{k+2},v_{k+3},v_{k+4},v_{k+5}]^T$, $\bar{v}^- = [v_{k-5},v_{k-4},v_{k-3},v_{k-2},v_{k-1},v_k]^T$, and $f_{LSTM}(\cdot)$ is an LSTM network.
The network is composed of an LSTM layer and a fully connected layer, where the LSTM layer is $6 \times 128$ and the fully connected layer is $128 \times 5$.
The LSTM network is trained using about six thousand vehicle trajectories from NGSIM data.

(S1) Zero-mean stochastic acceleration predictor.
\begin{equation}
a_{k+1} \sim \mathcal{N}(0,\sigma^2),
\end{equation}
where $\sigma$ is a constant, $\mathcal{N}$ stands for normal distribution.

(S2) Stochastic linear decay acceleration predictor.
\begin{equation}
\begin{aligned}
	\mu_{k+1} &= a_0 - (k+1)\frac{a_0}{\gamma}, \\
	a_{k+1} &\sim \mathcal{N}(\mu_{k+1},\sigma^2),
\end{aligned}
\end{equation}
where $\sigma$ and $\gamma$ are constants.

In simulation, 22 predictors are created based on the above six types. 
Three different linear decay acceleration predictors D2-a, D2-b and D2-c are used, where $\gamma = 3, 4$, and $5$ respectively.
Three different exponential decay acceleration predictors D3-a, D3-b and D3-c are used, where $\lambda = e^{-1}, e^{-2}$, and $e^{-3}$ respectively.
Seven different exponential decay acceleration predictors S1-a, S1-b, ..., S1-g are used, where $\sigma = 0.1, 0.2, 0.4, 0.6, 0.8, 1.0$, and $1.2$, respectively.
Seven different exponential decay acceleration predictors S2-a, S2-b, ..., S2-g are used, where $\sigma = 0.1, 0.2, 0.4, 0.6, 0.8, 1.0$, and $1.2$, respectively.
For all S2 predictors, $\gamma$ is set to 5 to match the best D2 predictor.


\begin{figure}[htbp]
\centerline{\includegraphics[width=\columnwidth]{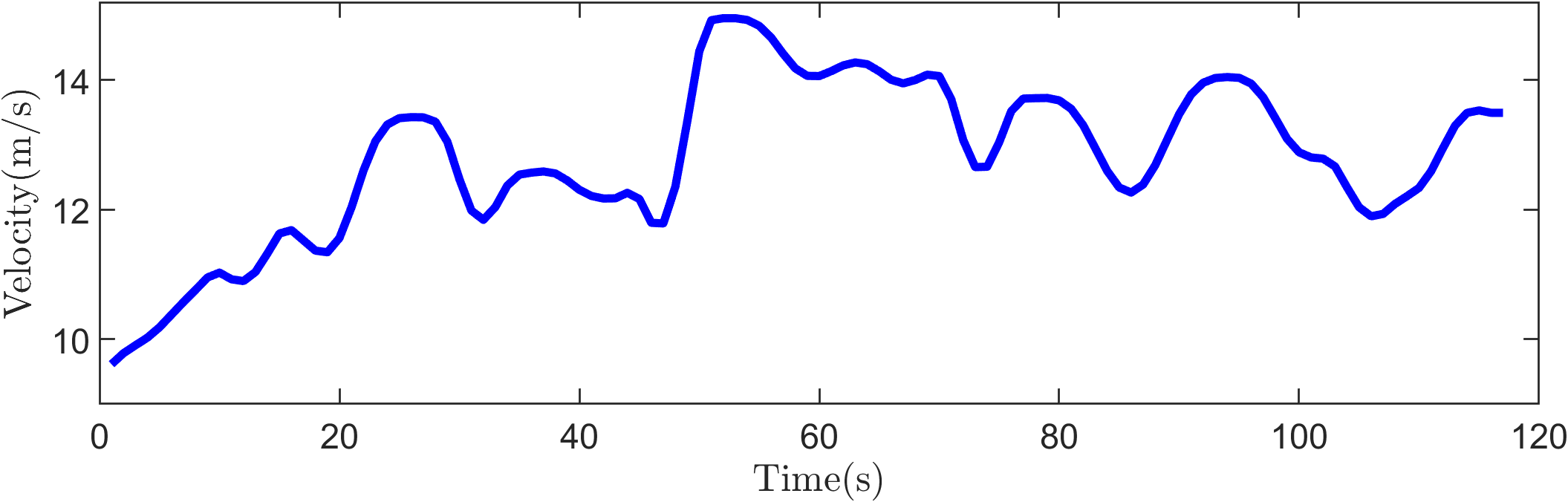}}
\caption{The driving data used in simulation}
\label{fig.driving_cycle}
\end{figure}

We run simulation over a 112-second driving data from NGSIM, as shown in Fig. \ref{fig.driving_cycle}.
The predictive optimal control is applied in a receding horizon fashion as the Type III problem.
The energy management strategy runs at 1 Hz, and the prediction horizon is 5 steps.

We use dynamic programming to approximately solve the optimal control problems after each prediction.
For stochastic predictors, the continuous probability distributions are discretized first.
Each normally-distributed random variable is approximated by a discrete random variable with 5 possible values, before dynamic programming is applied.

\begin{figure}[htbp]
\centerline{\includegraphics[width=\columnwidth]{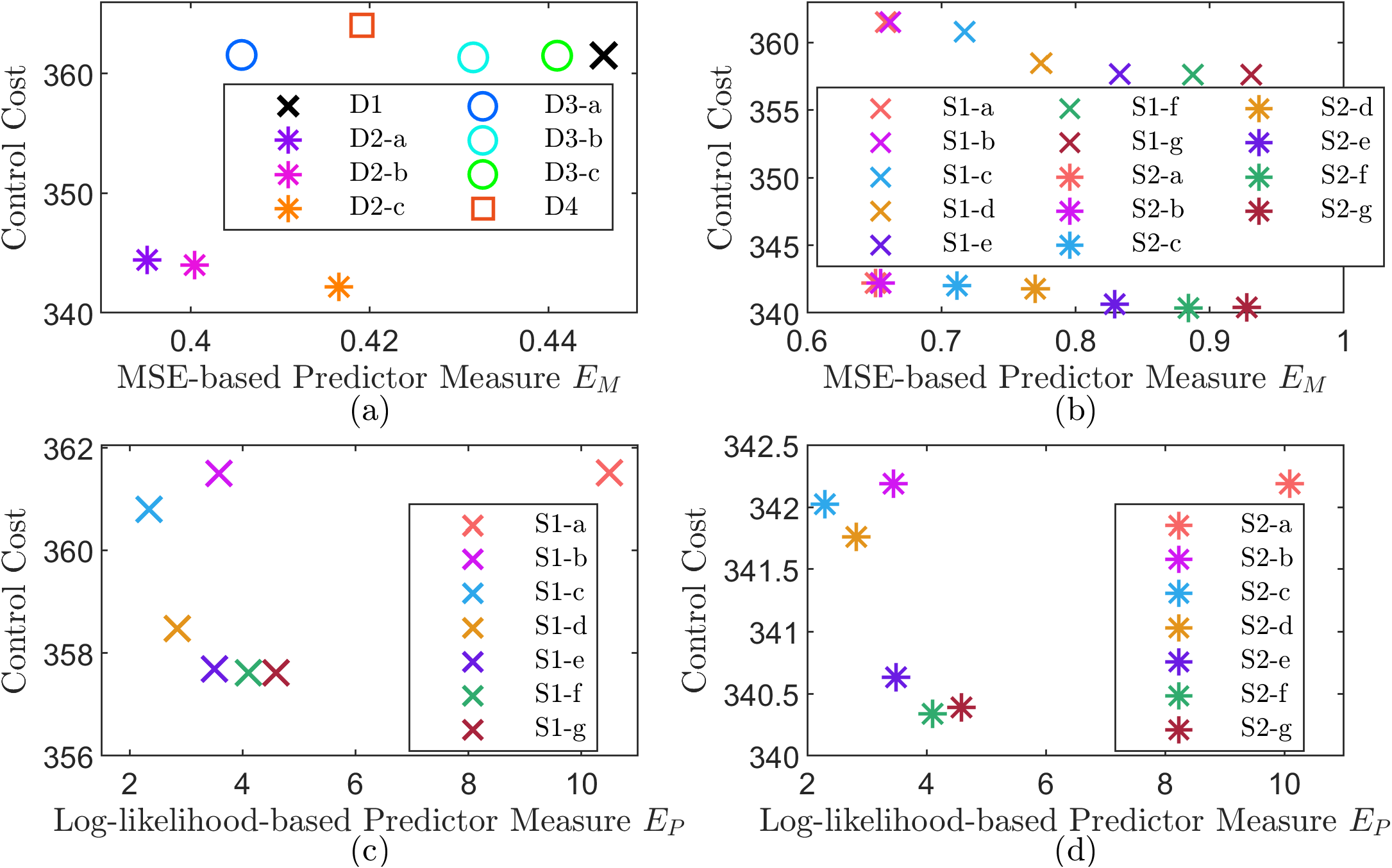}}
\caption{Automotive optimal control simulation results of prediction errors and control performance. (a) shows the MSE-based predictor measure and control cost of all deterministic predictors. (b) shows the MSE-based predictor measure and control cost of all stochastic predictors. (c) shows the log-likelihood-based predictor measure and control cost of all S1 stochastic predictors. (d) shows the log-likelihood-based predictor measure and control cost of all S2 stochastic predictors.}
\label{fig.result}
\end{figure}

The simulation results are shown in Fig. \ref{fig.result}.
In this Type III problem, the MSE and log-likelihood is defined as the average of multiple predictions.
All four plots in Fig. \ref{fig.result} show similar patterns to Fig. \ref{fig.cost_error} (b).
Therefore, when using the MSE or the log-likelihood as the predictor error measure, a better predictor does not necessarily mean better optimal control performance.
Comparing the deterministic predictor with its corresponding stochastic predictors (D1 vs S1, and D2 vs S2), it can be seen that in general the MSE-based predictor measures are larger for stochastic predictors.
However, the stochastic predictors may lead to lower control cost.
This implies that we should be cautious when directly comparing stochastic predictors with deterministic predictors.
It is still suggested that the predictors should not be evaluated alone, but with the optimal control task performance.


\section{Conclusions}
\label{sec:conclusions}

In this paper, an analysis framework for predictive optimal control is presented. 
An environment model which generates the to-be-predicted signal is included in the framework.
The truth of the to-be-predicted signal is properly defined with a hidden prediction state describing the current and future uncertainties in the environment. 
We use the proposed analysis framework to rethink the predictor evaluation problem.
It is shown that improving the predictor using a general performance measure may not guarantee the improvement in control performance.
It is suggested that for a general predictive control problem, the predictor should be evaluated along with the control system performance.

\section*{Acknowledgement}
This work was supported by the National Natural Science Foundation of China under Grant 52188102 and Grant 52272416.

\bibliography{poc}

\begin{thebibliography}{10}
\expandafter\ifx\csname url\endcsname\relax
  \def\url#1{\texttt{#1}}\fi
\expandafter\ifx\csname urlprefix\endcsname\relax\def\urlprefix{URL }\fi
\expandafter\ifx\csname href\endcsname\relax
  \def\href#1#2{#2} \def\path#1{#1}\fi

\bibitem{zengParallelHybridElectric2015a}
X.~Zeng, J.~Wang, A {{Parallel Hybrid Electric Vehicle Energy Management
  Strategy Using Stochastic Model Predictive Control With Road Grade Preview}},
  IEEE Transactions on Control Systems Technology 23~(6) (2015) 2416--2423.
\newblock \href {https://doi.org/10/f7vxn2} {\path{doi:10/f7vxn2}}.

\bibitem{wanProbabilisticAnticipationControl2019}
N.~Wan, C.~Zhang, A.~Vahidi, Probabilistic {{Anticipation}} and {{Control}} in
  {{Autonomous Car Following}}, IEEE Transactions on Control Systems Technology
  27~(1) (2019) 9.
\newblock \href {https://doi.org/10.1109/TCST.2017.2762288}
  {\path{doi:10.1109/TCST.2017.2762288}}.

\bibitem{YeAsurveyofmodel2019}
B.-L. Ye, W.~Wu, K.~Ruan, L.~Li, T.~Chen, H.~Gao, Y.~Chen, A survey of model
  predictive control methods for traffic signal control, IEEE/CAA Journal of
  Automatica Sinica 6~(3) (2019) 623--640.
\newblock \href {https://doi.org/10.1109/JAS.2019.1911471}
  {\path{doi:10.1109/JAS.2019.1911471}}.

\bibitem{fridovich-keilConfidenceawareMotionPrediction2020}
D.~Fridovich-Keil, A.~Bajcsy, J.~F. Fisac, S.~L. Herbert, S.~Wang, A.~D.
  Dragan, C.~J. Tomlin, Confidence-aware motion prediction for real-time
  collision avoidance, The International Journal of Robotics Research 39~(2-3)
  (2020) 250--265.
\newblock \href {https://doi.org/10.1177/0278364919859436}
  {\path{doi:10.1177/0278364919859436}}.

\bibitem{drgonaAllYouNeed2020}
J.~Drgo{\v{n}}a, J.~Arroyo, I.~C. Figueroa, D.~Blum, K.~Arendt, D.~Kim, E.~P.
  Oll{\'{e}}, J.~Oravec, M.~Wetter, D.~L. Vrabie, L.~Helsen, All you need to
  know about model predictive control for buildings, Annual Reviews in Control
  50 (2020) 190--232.
\newblock \href
  {https://doi.org/https://doi.org/10.1016/j.arcontrol.2020.09.001}
  {\path{doi:https://doi.org/10.1016/j.arcontrol.2020.09.001}}.

\bibitem{9922564}
C.~Tang, Y.~Liu, H.~Xiao, L.~Xiong, Integrated decision making and planning
  framework for autonomous vehicle considering uncertain prediction of
  surrounding vehicles, in: 2022 IEEE 25th International Conference on
  Intelligent Transportation Systems (ITSC), 2022, pp. 3867--3872.
\newblock \href {https://doi.org/10.1109/ITSC55140.2022.9922564}
  {\path{doi:10.1109/ITSC55140.2022.9922564}}.

\bibitem{8248668}
C.~Hubmann, J.~Schulz, M.~Becker, D.~Althoff, C.~Stiller, Automated driving in
  uncertain environments: Planning with interaction and uncertain maneuver
  prediction, IEEE Transactions on Intelligent Vehicles 3~(1) (2018) 5--17.
\newblock \href {https://doi.org/10.1109/TIV.2017.2788208}
  {\path{doi:10.1109/TIV.2017.2788208}}.

\bibitem{509370}
M.~Tomizuka, Model based prediction, preview and robust controls in motion
  control systems, in: Proceedings of 4th IEEE International Workshop on
  Advanced Motion Control - AMC '96 - MIE, Vol.~1, 1996, pp. 1--6 vol.1.
\newblock \href {https://doi.org/10.1109/AMC.1996.509370}
  {\path{doi:10.1109/AMC.1996.509370}}.

\bibitem{yuPowerPredictionsOnline2020}
C.~Yu, G.~Shi, S.-J. Chung, Y.~Yue, A.~Wierman,
  \href{http://arxiv.org/abs/2006.07569}{Competitive control with delayed
  imperfect information}, in: 2022 American Control Conference ({ACC}), {IEEE},
  2022.
\newblock \href {http://arxiv.org/abs/2006.07569} {\path{arXiv:2006.07569}},
  \href {https://doi.org/10.23919/acc53348.2022.9867421}
  {\path{doi:10.23919/acc53348.2022.9867421}}.
\newline\urlprefix\url{http://arxiv.org/abs/2006.07569}

\bibitem{mayneConstrainedModelPredictive2000}
D.~Q. Mayne, J.~B. Rawlings, C.~V. Rao, P.~O.~M. Scokaert, Constrained model
  predictive control: {{Stability}} and optimality, Automatica 36~(6) (2000)
  26.
\newblock \href {https://doi.org/10/brxgpc} {\path{doi:10/brxgpc}}.

\bibitem{ZHANG2023111101}
Y.~Zhang, C.~Edwards, M.~Belmont, G.~Li,
  \href{https://www.sciencedirect.com/science/article/pii/S0005109823002613}{Robust
  model predictive control for constrained linear system based on a sliding
  mode disturbance observer}, Automatica 154 (2023) 111101.
\newblock \href
  {https://doi.org/https://doi.org/10.1016/j.automatica.2023.111101}
  {\path{doi:https://doi.org/10.1016/j.automatica.2023.111101}}.
\newline\urlprefix\url{https://www.sciencedirect.com/science/article/pii/S0005109823002613}

\bibitem{vallon_data-driven_2022}
C.~S. Vallon, F.~Borrelli,
  \href{https://ieeexplore.ieee.org/document/9675285/}{Data-{Driven}
  {Strategies} for {Hierarchical} {Predictive} {Control} in {Unknown}
  {Environments}}, IEEE Transactions on Automation Science and Engineering
  19~(3) (2022) 1434--1445.
\newblock \href {https://doi.org/10.1109/TASE.2021.3137769}
  {\path{doi:10.1109/TASE.2021.3137769}}.
\newline\urlprefix\url{https://ieeexplore.ieee.org/document/9675285/}

\bibitem{DEY2023100255}
S.~Dey, T.~Marzullo, X.~Zhang, G.~Henze,
  \href{https://www.sciencedirect.com/science/article/pii/S2666546823000277}{Reinforcement
  learning building control approach harnessing imitation learning}, Energy and
  AI 14 (2023) 100255.
\newblock \href {https://doi.org/https://doi.org/10.1016/j.egyai.2023.100255}
  {\path{doi:https://doi.org/10.1016/j.egyai.2023.100255}}.
\newline\urlprefix\url{https://www.sciencedirect.com/science/article/pii/S2666546823000277}

\bibitem{MUGNINI20232169}
A.~Mugnini, F.~Ferracuti, M.~Lorenzetti, G.~Comodi, A.~Arteconi,
  \href{https://www.sciencedirect.com/science/article/pii/S2352484723000136}{Day-ahead
  optimal scheduling of smart electric storage heaters: A real quantification
  of uncertainty factors}, Energy Reports 9 (2023) 2169--2184.
\newblock \href {https://doi.org/https://doi.org/10.1016/j.egyr.2023.01.013}
  {\path{doi:https://doi.org/10.1016/j.egyr.2023.01.013}}.
\newline\urlprefix\url{https://www.sciencedirect.com/science/article/pii/S2352484723000136}

\bibitem{PANAGIOTELIS2023693}
A.~Panagiotelis, P.~Gamakumara, G.~Athanasopoulos, R.~J. Hyndman,
  \href{https://www.sciencedirect.com/science/article/pii/S0377221722006087}{Probabilistic
  forecast reconciliation: Properties, evaluation and score optimisation},
  European Journal of Operational Research 306~(2) (2023) 693--706.
\newblock \href {https://doi.org/https://doi.org/10.1016/j.ejor.2022.07.040}
  {\path{doi:https://doi.org/10.1016/j.ejor.2022.07.040}}.
\newline\urlprefix\url{https://www.sciencedirect.com/science/article/pii/S0377221722006087}

\bibitem{hu_multihorizon_2022}
Q.~Hu, M.~R. Amini, I.~Kolmanovsky, J.~Sun, A.~Wiese, J.~B. Seeds,
  \href{https://ieeexplore.ieee.org/document/9478061/}{Multihorizon {Model}
  {Predictive} {Control}: {An} {Application} to {Integrated} {Power} and
  {Thermal} {Management} of {Connected} {Hybrid} {Electric} {Vehicles}}, IEEE
  Transactions on Control Systems Technology 30~(3) (2022) 1052--1064.
\newblock \href {https://doi.org/10.1109/TCST.2021.3091887}
  {\path{doi:10.1109/TCST.2021.3091887}}.
\newline\urlprefix\url{https://ieeexplore.ieee.org/document/9478061/}

\bibitem{palmerStochasticWeatherClimate2019}
T.~N. Palmer, Stochastic weather and climate models, Nature Reviews Physics
  1~(7) (2019) 463--471.
\newblock \href {https://doi.org/10.1038/s42254-019-0062-2}
  {\path{doi:10.1038/s42254-019-0062-2}}.

\bibitem{DensityForecastingSurvey}
A.~S. Tay, K.~F. Wallis, C.~C. Al, Density forecasting: A survey, Journal of
  Forecasting 19~(4) (2000) 235--254.
\newblock \href
  {https://doi.org/10.1002/1099-131x(200007)19:4<235::aid-for772>3.0.co;2-l}
  {\path{doi:10.1002/1099-131x(200007)19:4<235::aid-for772>3.0.co;2-l}}.

\bibitem{gneitingProbabilisticForecasting2014}
T.~Gneiting, M.~Katzfuss, Probabilistic {{Forecasting}}, Annual Review of
  Statistics and Its Application 1~(1) (2014) 125--151.
\newblock \href {https://doi.org/10.1146/annurev-statistics-062713-085831}
  {\path{doi:10.1146/annurev-statistics-062713-085831}}.

\bibitem{denoeuxDecisionmakingBeliefFunctions2019}
T.~Denœux, Decision-making with belief functions: {{A}} review, International
  Journal of Approximate Reasoning 109 (2019) 87--110.
\newblock \href {https://doi.org/10.1016/j.ijar.2019.03.009}
  {\path{doi:10.1016/j.ijar.2019.03.009}}.

\bibitem{petropoulosForecastingTheoryPractice2022}
F.~Petropoulos, D.~Apiletti, V.~Assimakopoulos, M.~Z. Babai, D.~K. Barrow,
  S.~B. Taieb, C.~Bergmeir, R.~J. Bessa, J.~Bijak, J.~E. Boylan, J.~Browell,
  C.~Carnevale, J.~L. Castle, P.~Cirillo, M.~P. Clements, C.~Cordeiro, F.~L.~C.
  Oliveira, S.~De~Baets, A.~Dokumentov, J.~Ellison, P.~Fiszeder, P.~H. Franses,
  D.~T. Frazier, M.~Gilliland, M.~S. Gönül, P.~Goodwin, L.~Grossi,
  Y.~Grushka-Cockayne, M.~Guidolin, M.~Guidolin, U.~Gunter, X.~Guo, R.~Guseo,
  N.~Harvey, D.~F. Hendry, R.~Hollyman, T.~Januschowski, J.~Jeon, V.~R.~R.
  Jose, Y.~Kang, A.~B. Koehler, S.~Kolassa, N.~Kourentzes, S.~Leva, F.~Li,
  K.~Litsiou, S.~Makridakis, G.~M. Martin, A.~B. Martinez, S.~Meeran, T.~Modis,
  K.~Nikolopoulos, D.~Önkal, A.~Paccagnini, A.~Panagiotelis, I.~Panapakidis,
  J.~M. Pavía, M.~Pedio, D.~J. Pedregal, P.~Pinson, P.~Ramos, D.~E. Rapach,
  J.~J. Reade, B.~Rostami-Tabar, M.~Rubaszek, G.~Sermpinis, H.~L. Shang,
  E.~Spiliotis, A.~A. Syntetos, P.~D. Talagala, T.~S. Talagala, L.~Tashman,
  D.~Thomakos, T.~Thorarinsdottir, E.~Todini, J.~R.~T. Arenas, X.~Wang, R.~L.
  Winkler, A.~Yusupova, F.~Ziel, Forecasting: Theory and practice,
  International Journal of Forecasting 38~(3) (2022) S0169207021001758.
\newblock \href {http://arxiv.org/abs/2012.03854} {\path{arXiv:2012.03854}},
  \href {https://doi.org/10.1016/j.ijforecast.2021.11.001}
  {\path{doi:10.1016/j.ijforecast.2021.11.001}}.

\bibitem{StochasticModelPredictive2016a}
A.~Mesbah, Stochastic model predictive control: An overview and perspectives
  for future research, IEEE Control Systems Magazine 36~(6) (2016) 30--44.
\newblock \href {https://doi.org/10.1109/MCS.2016.2602087}
  {\path{doi:10.1109/MCS.2016.2602087}}.

\bibitem{rosoliaDataDrivenPredictiveControl2018}
U.~Rosolia, X.~Zhang, F.~Borrelli, Data-{{Driven Predictive Control}} for
  {{Autonomous Systems}}, Annual Review of Control, Robotics, and Autonomous
  Systems 1~(1) (2018) 259--286.
\newblock \href {https://doi.org/10/gg4z5d} {\path{doi:10/gg4z5d}}.

\bibitem{mandiSmartPredictandOptimizeHard2020}
J.~Mandi, E.~Demirovi?, P.~J. Stuckey, T.~Guns, Smart {{Predict-and-Optimize}}
  for {{Hard Combinatorial Optimization Problems}}, Proceedings of the AAAI
  Conference on Artificial Intelligence 34~(02) (2020) 1603--1610.
\newblock \href {https://doi.org/10.1609/aaai.v34i02.5521}
  {\path{doi:10.1609/aaai.v34i02.5521}}.

\bibitem{elmachtoubSmartPredictThen2020a}
A.~N. Elmachtoub, P.~Grigas, \href{http://arxiv.org/abs/1710.08005}{Smart
  "{{Predict}}, then {{Optimize}}"}, Management Science 68~(1) (2020) 9--26.
\newblock \href {http://arxiv.org/abs/1710.08005} {\path{arXiv:1710.08005}},
  \href {https://doi.org/10.1287/mnsc.2020.3922}
  {\path{doi:10.1287/mnsc.2020.3922}}.
\newline\urlprefix\url{http://arxiv.org/abs/1710.08005}

\bibitem{wilderMeldingDataDecisionsPipeline2019}
B.~Wilder, B.~Dilkina, M.~Tambe, Melding the {{Data-Decisions Pipeline}}:
  {{Decision-Focused Learning}} for {{Combinatorial Optimization}}, Proceedings
  of the AAAI Conference on Artificial Intelligence 33~(01) (2019) 1658--1665.
\newblock \href {https://doi.org/10.1609/aaai.v33i01.33011658}
  {\path{doi:10.1609/aaai.v33i01.33011658}}.

\bibitem{bayerRobustEconomicModel2016}
F.~A. Bayer, M.~Lorenzen, M.~A. Müller, F.~Allgöwer, Robust economic {{Model
  Predictive Control}} using stochastic information, Automatica 74 (2016)
  151--161.
\newblock \href {https://doi.org/10.1016/j.automatica.2016.08.008}
  {\path{doi:10.1016/j.automatica.2016.08.008}}.

\bibitem{jingPredictiveOptimalControl2021}
R.~Jing, X.~Zeng, Predictive {{Optimal Control}} with {{Data-Based Disturbance
  Scenario Tree Approximation}}, in: 2021 {{American Control Conference}}
  ({{ACC}}), {IEEE}, {New Orleans, LA, USA}, 2021, pp. 992--997.
\newblock \href {https://doi.org/10.23919/ACC50511.2021.9483341}
  {\path{doi:10.23919/ACC50511.2021.9483341}}.

\bibitem{TransportationFederalHighwayAdministration2016}
U.~D. of~Transportation Federal Highway~Administration, Next generation
  simulation ({NGSIM}) vehicle trajectories and supporting data, provided by
  ITS DataHub through Data.transportation.gov. Accessed 2023-02-01 from
  http://doi.org/10.21949/1504477" (2016).
\newblock \href {https://doi.org/http://doi.org/10.21949/1504477}
  {\path{doi:http://doi.org/10.21949/1504477}}.

\bibitem{zengOptimizingEnergyManagement2017}
X.~Zeng, J.~Wang, Optimizing the energy management strategy for plug-in hybrid
  electric vehicles with multiple frequent routes, {IEEE} Transactions on
  Control Systems Technology 27~(1) (2019) 394--400.
\newblock \href {https://doi.org/10.1109/tcst.2017.2768042}
  {\path{doi:10.1109/tcst.2017.2768042}}.

\end{thebibliography}
\bibliographystyle{elsarticle-num}

\end{document}